\documentclass[11pt]{article}

\usepackage[final]{acl}

\usepackage{times}
\usepackage{latexsym}
\usepackage{multirow}
\usepackage[T1]{fontenc}

\usepackage[utf8]{inputenc}

\usepackage{microtype}

\usepackage{inconsolata}

\usepackage{tikz}

\usepackage{booktabs}
\usepackage{graphicx}
\usepackage{pifont}
\usepackage{makecell} 
\usepackage{xcolor}
\usepackage{colortbl}
\usepackage{amsmath}
\usepackage{amsthm}
\usepackage{bm}
\usepackage[most]{tcolorbox}
\usepackage{enumitem}
\usepackage{placeins}
\usepackage{pgfplots} 
\pgfplotsset{compat=1.18}
\usepackage{titletoc}
\usepackage{CJKutf8} 
\usepackage{subcaption}

\newcommand{\cmark}{\ding{51}} 
\newcommand{\xmark}{\ding{55}} 

\theoremstyle{plain}

\title{
UniFinEval: Towards Unified Evaluation of Financial Multimodal Models across Text, Images  and Videos}

\vspace{5mm}

\author{
\textbf{Zhi Yang\textsuperscript{1}}\footnotemark[1],
 \textbf{Lingfeng Zeng\textsuperscript{1}}\thanks{Equal contribution.},
 \textbf{Fangqi Lou\textsuperscript{1}}\footnotemark[1],
 \textbf{Qi Qi\textsuperscript{1}},
 \textbf{Wei Zhang\textsuperscript{1}},
\\
\textbf{Zhenyu Wu\textsuperscript{1}},
 \textbf{Zhenxiong Yu\textsuperscript{1}},
 \textbf{Jun Han\textsuperscript{1}},
 \textbf{Zhiheng Jin\textsuperscript{1}},
 \textbf{Lejie Zhang\textsuperscript{1}},
\\
 \textbf{Xiaoming Huang\textsuperscript{2}},
 \textbf{Xiaolong Liang\textsuperscript{2}},
 \textbf{Zheng Wei\textsuperscript{2}},
 \textbf{Junbo Zou\textsuperscript{3}},
 \textbf{Dongpo Cheng\textsuperscript{1}},
\\
\textbf{Zhaowei Liu\textsuperscript{1}},
 \textbf{Xin Guo\textsuperscript{1}},
 \textbf{Rongjunchen Zhang\textsuperscript{4}}\footnotemark[2],
 \textbf{Liwen Zhang\textsuperscript{1}}\thanks{Corresponding authors.
\href{mailto:zhang.liwen@shufe.edu.cn}{zhang.liwen@shufe.edu.cn}, \href{mailto:zhangrongjunchen@myhexin.com}{zhangrongjunchen@myhexin.com}.}\vspace{1mm}
\\
 \textsuperscript{1}SUFE,\,
 \textsuperscript{2}Tencent,\,
  \textsuperscript{3}Gatech,\,
    \textsuperscript{4}HiThink Research \vspace{2mm}
\\
 {
   \href{zhang.liwen@shufe.edu.cn}{zhang.liwen@shufe.edu.cn}
 }
}

\begin{document}
\maketitle

\begin{abstract}
Multimodal large language models are playing an increasingly significant role in empowering the financial domain, however, the challenges they face, such as multimodal and high-density information and cross-modal multi-hop reasoning, go beyond the evaluation scope of existing multimodal benchmarks. 
To address this gap, we propose \textbf{UniFinEval}, the first unified multimodal benchmark designed for high-information-density financial environments, covering text, images, and videos. UniFinEval systematically constructs five core financial scenarios grounded in real-world financial systems: Financial Statement Auditing, Company Fundamental Reasoning, Industry Trend Insights, Financial Risk Sensing, and Asset Allocation Analysis. We manually construct a high-quality dataset consisting of 3,767 question-answer pairs in both chinese and english and systematically evaluate 10 mainstream MLLMs under Zero-Shot and CoT settings. Results show that Gemini-3-pro-preview achieves the best overall performance, yet still exhibits a substantial gap compared to financial experts. Further error analysis reveals systematic deficiencies in current models.
UniFinEval aims to provide a systematic assessment of MLLMs’ capabilities in fine-grained, high–information-density financial environments, thereby enhancing the robustness of MLLMs applications in real-world financial scenarios. Data and code are available at \url{https://github.com/aifinlab/UniFinEval}.
\end{abstract}

\begin{figure}[ht] 
    \centering
    \includegraphics[width=0.5\textwidth]{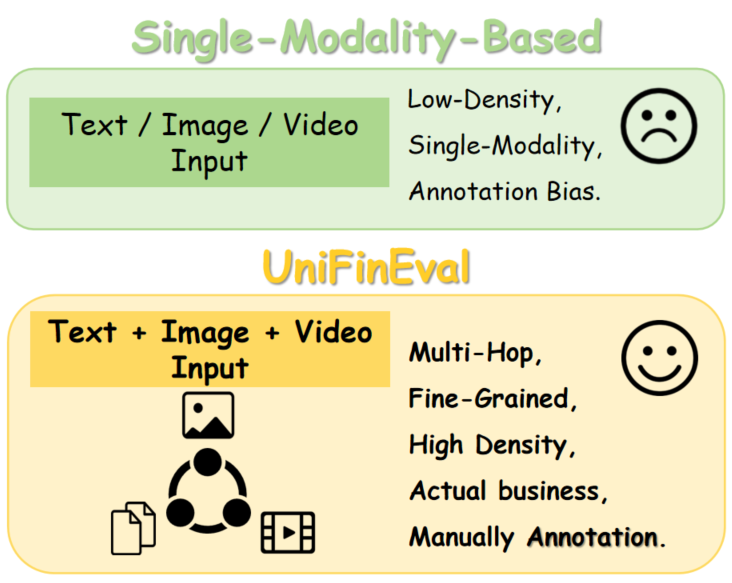} 
    \caption{UniFinEval is manually constructed and supports full-modality inputs including text, images, and videos. It is equipped with cross-modal reasoning capabilities and features high information density while closely aligning with real financial business practices. }
    \label{fig:compare}
    \vspace{-2pt}
\end{figure}

\newpage

\section{Introduction}

In the current era of rapid multimodal large language models(MLLMs) development and explosive information growth, the demand for utilizing large language models(LLMs) to process information across various fields is becoming increasingly urgent~\cite{rahman2025llm,li2025survey,zhang2025data,dennstadt2025comparative,xia2024sportu,10.1145/3746027.3754557}. In real-world financial scenarios, models are often required to simulate the role of analysts, simultaneously processing voluminous financial and research reports, understanding implicit correspondences between charts and text, and conducting continuous analysis by incorporating financial analysis videos.
However, there is a significant misalignment between existing multimodal financial benchmarks and these real-world demands. As illustrated in Figure~\ref{fig:compare} these limitations are primarily manifested in the following aspects: On one hand, existing benchmarks are limited to single-modality evaluations, such as FinVQA~\cite{chen2021finqa}, Fin-Fact~\cite{rangapur2025fin}, and MMMU~\cite{yue2024mmmu}, and have not yet extended to broader multi-modal tasks. This modality-restricted approach leads to inconsistencies in characterizing model capabilities and limits the reference value of evaluation results for real-world financial applications.  On the other hand, the high-density and noisy information environment of real financial markets places higher demands on models' fine-grained analysis capabilities. However, existing studies mostly use LLMs to construct datasets based on simplified or truncated data~\cite{luo2025finmme,liu-etal-2025-visfineval}. This not only makes it difficult to evaluate the comprehensive performance of models in high-information-density environments but also introduces potential annotation errors, thereby affecting the reliability of the evaluation conclusions. These bias makes it difficult for a model's performance on benchmarks to reflect its capability boundaries in real financial operations. Consequently, this may lead to unstable analysis conclusions, exposure to compliance risks, and even severe financial losses in practical applications~\cite{lee2024survey}.

\begin{figure*}[ht]   
    \centering
    \includegraphics[width=\textwidth]{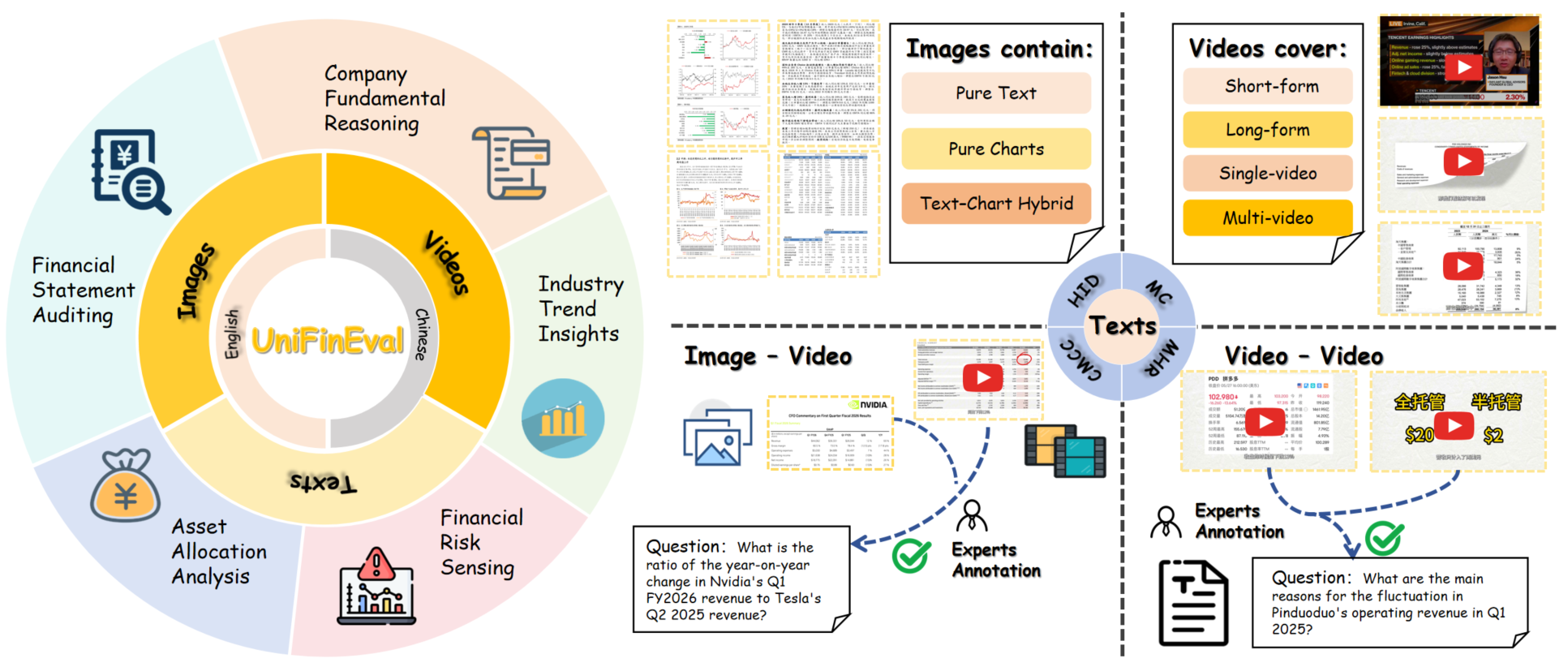}
    \caption{UniFinEval covers five major financial scenarios and constructs datasets spanning text, images, videos, as well as multiple cross-modal combinations. It features high-information-density and manually construct data, together with dedicated designs for cross-modal consistency checking and multi-Hop reasoning, providing comprehensive support for MLLMs evaluation in financial domains. }
    \label{fig:frame}
    \vspace{-2pt}
\end{figure*}

To fill these gaps, we propose UniFinEval, the first financial unified benchmark designed for high-information-density financial environments, integrating text, images, and videos modalities. UniFinEval grounded in real-world financial business workflows and systematically covers five core scenarios: Financial Statement Auditing, Company Fundamental Reasoning, Industry Trend Insights, Financial Risk Sensing, and Asset Allocation Analysis, completely characterizing complete cognitive loop from multimodal perception and cross-modal alignment to high-level decision making. All samples in this benchmark were manually constructed by financial experts, constructing a dataset of 3,767 high-quality Q\&A pairs aimed at assessing the capability boundaries of MLLMs in real financial scenarios. The main contributions of this paper are summarized as follows:

(1)~We propose UniFinEval, the first multimodal unified financial evaluation benchmark. By constructing \textbf{manually curated} question-answer pairs that deeply integrate text, images, and videos, we provide a unified paradigm for analyzing the profound capabilities of MLLMs within the financial domain. The structure is illustrated in Figure~\ref{fig:frame}.

(2)~We meticulously design multi-hop reasoning questions from the unique perspective of multi-source information fusion within the financial domain, thereby posing a rigorous challenge to the actual performance of MLLMs.

(3)~We conducted a targeted error analysis aimed at optimizing the capabilities of MLLMs within high-information-density complex application scenarios in the financial domain, providing practical and feasible directions for improvement.

The organization of this paper is as follows: Section~\ref{sec2} reviews the research on financial MLLMs and related multimodal benchmarks. Section~\ref{sec3} details the construction process of UniFinEval, including data collection, question design, and quality control processes. Sections~\ref{sec4} and~\ref{sec5} present our experimental settings and results under different difficulty levels, followed by error analysis. Finally, Section~\ref{sec6} summarizes the work of this paper and discusses potential future directions in the field of multimodal financial intelligence.

\section{Related Work}
\label{sec2}

\subsection{Financial Applications Analysis}
Benefiting from breakthrough advancements in general Large Language Models (LLMs), research on intelligence in the financial vertical domain has made significant progress. Early work mainly focused on tasks such as financial sentiment analysis~\cite{delgadillo2024finsosent,kim2024financial,iacovides2024finllama}, financial time series analysis~\cite{li2024alphafin,li2024finreport,wang2024quantagent,mai2024stockgpt,cao2024risklabs}, financial text understanding~\cite{masry2024longfin,wilson2024fin2sum}, and decision support~\cite{yu2024fincon,yu2024finmem,liu2025fin,xiao2025trading}. However, real-world financial information inherently possesses highly multimodal characteristics. To further enhance the deployment capabilities of models in real scenarios, the focus of related research has gradually expanded from a single text modality to a multimodal perspective. FinVis-GPT~\cite{wang2023finvis}, Finzero~\cite{wang2025finzero}, and FinTral~\cite{bhatia2024fintral} achieve deep analysis of financial charts. FinAgent~\cite{zhang2024finagent} created a multimodal agent capable of autonomous trading decisions, MM-DREX ~\cite{chen2025mm}utilizes multimodal information to implement dynamic routing for expert trading; and AT-FinGPT ~\cite{liu2025fingpt} utilizes voice features to assist in financial risk prediction.

\subsection{Multimodal Financial Benchmark Analysis}
Compared to the rapid expansion of model capabilities, existing financial multimodal evaluation systems appear significantly lagged. On one hand, traditional financial benchmarks are confined to the text modality ~\cite{zhu2024benchmarking,nie2024cfinbench,zhao2024financemath,wang2024doctabqa,reddy2024docfinqa,chen2024fintextqa,chen2025mtbench,guo-etal-2025-fineval,liu2025findabench,li-etal-2025-investorbench,matlin-etal-2025-financial}, while general multimodal benchmarks lack domain-specific knowledge depth ~\cite{liu2024mmbench,li2024seed,zhang2024mme,yu2024mm}; neither can accurately measure the financial multimodal capabilities of financial large models. On the other hand, although current financial multimodal evaluation benchmarks have made progress in specific tasks, they still exhibit significant limitations. Specifically, benchmarks such as MME-Finance ~\cite{gan2024mme}, FinChart-Bench ~\cite{shu2025finchart}, CFBenchmark-MM ~\cite{li2025cfbenchmark}, and MultiFinBen ~\cite{peng2025multifinben} mainly focus on static chart understanding, lacking consideration for dynamic time-series information such as financial videos. Data for works like FinMR ~\cite{deng2025finmr}, FAMMA ~\cite{xue2024famma}, and XFinBench ~\cite{zhang2025xfinbench} are mostly derived from textbooks or standardized exams, creating a large gap with real financial environments full of noise and unstructured information. Works like FinMMR ~\cite{tang2025finmmr} and FinMultiTime ~\cite{xu2025finmultitime} utilize large model annotation paradigms, potentially introducing model bias and hallucination risks. Furthermore, VideoConviction ~\cite{galarnyk2025videoconviction} and Fincap ~\cite{sukhani2025fincap}, among the few benchmarks involving financial videos, neither evaluate financial business competence nor assess joint reasoning over video and text–image information.

\section{UniFinEval}
\label{sec3}

\subsection{Overview}
To evaluate the performance and failure modes of MLLMs in real-world financial business environments, we propose \textbf{UniFinEval}, a benchmark designed around complex and information-dense financial scenarios.
Guided by financial experts and grounded in authentic business practices, UniFinEval identifies five representative and critical financial scenarios:
Financial Statement Auditing (FSA), Company Fundamental Reasoning (CFR), Industry Trend Insights (ITI), Financial Risk Sensing (FRS), and Asset Allocation Analysis (AAA).

These scenarios cover a broad spectrum of financial reasoning tasks, ranging from fine-grained information verification and consistency checking to cross-firm and cross-cycle analysis, and ultimately high-level risk control and decision-making.
Such a progression reflects increasing requirements on models’ information perception, cross-modal alignment, multi-hop reasoning, and decision robustness. In addition, UniFinEval incorporates a series of task-specific perturbations aligned with real-world financial business settings into each scenario.
These perturbations enable the evaluation of model behavior under non-ideal visual and textual conditions commonly encountered in practical financial applications; details are provided in Appendix~\ref{sec:perturbation}.

\begin{figure*}[ht]   
    \centering
    \includegraphics[width=0.78\textwidth]{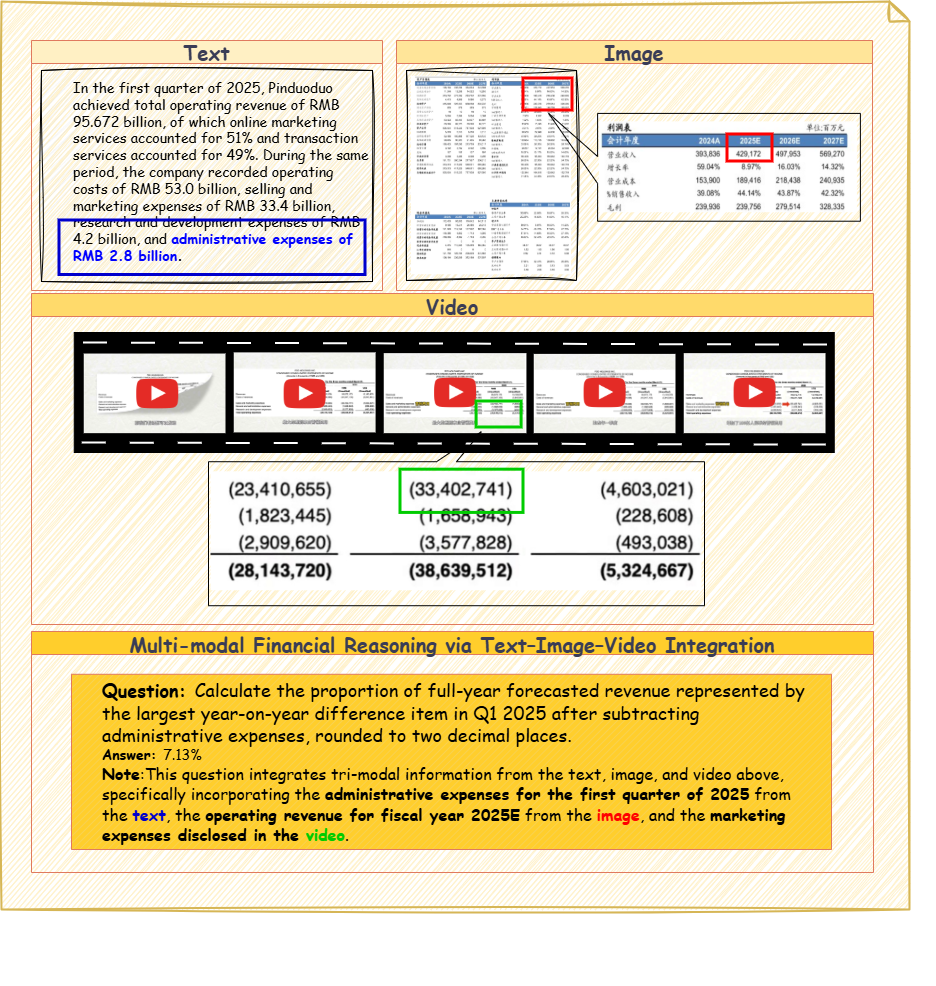}
    \vspace{-15pt}
    \caption{An example of a cross-modal multi-hop question in UniFinEval. The answer is derived from the acquisition and integration of key information from the presented text, images, and videos.}
    \label{fig:examgle}
\end{figure*}

Centered on these business scenarios, UniFinEval is guided throughout by financial experts to construct a data system featuring high information density and deep multi-modal fusion. The benchmark facilitates multi-hop integration across three foundational modalities—text, images, and video—by systematically incorporating complex cross-modal combinations, including text–image, text–video, image–video, and text–image–video, as illustrated in Figure~\ref{fig:examgle}. UniFinEval authentically replicates the parallel information structure of professional financial workflows. 
Through a rigorous multi-layer quality control process, we construct a dataset comprising 3,767 high-quality bilingual (Chinese and English) Q\&A pairs. A detailed breakdown of data distribution across the five financial scenarios is provided in Table~\ref{tab:Unifineval_dist} in the Appendix.
Regarding task design, the benchmark incorporates both \textit{single-turn} and \textit{multi-turn} Q\&A mechanisms and extensively integrates cross-modal multi-hop reasoning requirements. Such a design enables the assessment of the model's comprehensive capabilities in immediate understanding, cross-turn context integration, and complex logic chain construction, providing a solid foundation for systematically evaluating the performance of MLLMs in real financial environments.

\subsection{Scenario Construction}
The scenario construction of UniFinEval is not centered on isolated tasks or individual model capabilities, but is designed to align with real-world financial business processes.
In practical financial settings, performance on a single capability is often insufficient to support reliable deployment; instead, model stability and consistency across multiple business stages are critical considerations for real applications.
Motivated by this observation, UniFinEval establishes five interconnected and progressively hierarchical financial business scenarios.
As the scenarios advance, the object of analysis expands from localized information processing to global judgment, while data forms evolve from static inputs to dynamic multi-modal combinations.
Accordingly, the demands placed on the model increase across stages, and performance at each step provides insight into the challenges and feasibility of deploying MLLMs in realistic financial environments.

\textbf{Financial Statement Auditing: }
As the primary stage of financial analysis, this scenario focuses on verifying the accuracy and consistency of financial report information, with the goal of establishing a reliable data foundation for subsequent end-to-end analysis. To address the practical challenge faced by practitioners when verifying information in complex visual environments, we design high–information-density images that integrate textual content, charts, and their coupled representations. Authentic layouts and redundant information are deliberately preserved to reflect real-world business settings. Models are required to complete key information extraction and verification through both single-turn and multi-turn Q\&A, enabling the evaluation of their ability to precisely locate and validate critical financial facts under conditions of high information density, complex layouts, and redundant interference.

\textbf{Company Fundamental Reasoning: }
Following the FSA scenario, the CFR scenario focuses on analyzing corporate operating conditions and intrinsic value. To align with real-world practice in which practitioners examine financial reports alongside research reports, the data in this scenario continue to use text and charts as primary information carriers, while placing greater emphasis on financial variables and calculation bases distributed across heterogeneous modalities. The task design centers on the derivation of core financial indicators, requiring the model to extract relevant parameters from multi-source information and perform rigorous multi-step calculations. By incorporating high-difficulty cross-modal multi-hop reasoning, the scenario effectively differentiates basic information retrieval ability from deeper financial mathematical and logical reasoning capabilities.

\textbf{Industry Trend Insights: } 
The ITI scenario elevates the analytical perspective from individual enterprises to the industry level, focusing on cross-enterprise and cross-period analysis to evaluate the model’s ability to integrate multi-source information and infer industry logic. To reflect real-world demands for cross–data-source industry research, the data construction integrates multi-period financial reports, industry research reports, and macroeconomic data. A broader informational context is constructed using text and trend charts as primary carriers. The tasks center on industry trend assessment and cycle summarization, requiring complex cross-enterprise comparative reasoning and multi-hop inference. Through single-turn and multi-turn interactions, the scenario simulates how practitioners extract structured industry insights from fragmented information.

\textbf{Financial Risk Sensing: } 
The scenario focuses on the multi-dimensional identification and integration of potential risk signals, serving as a critical component in safeguarding investment decisions within financial analysis. It emphasizes evaluating the model’s robustness and risk perception capabilities under conditions of dynamic multi-modal information fusion. Since risk signals are often implicit and embedded in unstructured and time-varying information, the scenario introduces financial analysis videos, constructing a \emph{text–image–video} multi-modal setting. Dynamic viewpoints presented in videos are explicitly aligned with written quantitative data, restoring the multi-source cross-validation requirements characteristic of real-world risk analysis. The tasks center on risk signal identification and cross-modal consistency verification, incorporating cross-modal multi-hop reasoning. This design requires the model to jointly assess heterogeneous information sources and accurately capture latent downside risk signals.

\textbf{Asset Allocation Analysis:} 
As the final decision-making stage in financial workflows, Asset Allocation Analysis requires synthesizing insights from all preceding analytical stages to formulate executable strategies under multiple constraints. In this scenario, UniFinEval constructs the most complex input structure, integrating multi-modal data from all prior scenarios while further introducing mixed \emph{text–image–video} inputs and explicit real-world constraints, thereby closely approximating the information boundaries of realistic decision-making processes. The task design follows the complete asset allocation pipeline and primarily adopts multi-turn Q\&A interactions, requiring the model to iteratively integrate prior analytical results to produce logically consistent allocation strategies. This scenario ultimately evaluates the model’s comprehensive decision-making ability and its capacity to manage trade-offs under high information density and complex constraints.

\emph{Overall}, through the progressive construction of the five scenarios described above, UniFinEval establishes a comprehensive evaluation framework that spans from static high-density perception, to dynamic cross-modal reasoning, and ultimately to expert-level decision support. As the scenarios advance, data modalities become increasingly diverse and task structures grow more complex, enabling fine-grained and interpretable quantitative assessment of model capabilities across different cognitive levels. This framework provides a reliable reference for evaluating the practical applicability of MLLMs in real-world financial business environments.

\subsection{Quality Control}

Given the instability and potential bias of current multimodal models in complex financial tasks, as well as the financial industry’s stringent requirements for precision, interpretability, and logical consistency, UniFinEval adopts a fully expert-driven manual construction strategy for question generation and validation. To eliminate any form of model-induced bias, no LLMs are involved at any stage of Q\&A creation, ensuring that all evaluation tasks strictly adhere to real-world financial business logic and regulatory compliance standards.

The dataset construction is carried out by a dedicated team of ten senior financial experts, including PhD students from top-tier finance and economics institutions and experienced practitioners from leading securities firms and financial institutions. All experts hold professional certifications such as CFA or CPA, with an average of more than five years of frontline industry or research experience. This dual composition of academic and industry expertise ensures both theoretical rigor and practical relevance in scenario design.

Question construction follows a rigorous four-stage quality control pipeline. First, during data filtering, automated scripts combined with manual inspection are used to select multimodal financial materials with high information density and strong business relevance. Second, domain experts independently formulate questions and corresponding standard answers based on the curated data. Third, all annotations undergo cross-validation by other experts to identify logical inconsistencies, ambiguity, or deviations from realistic business practices. Finally, a unified audit is conducted to ensure that the resulting tasks faithfully reflect the complexity, decision constraints, and reasoning processes encountered in real financial environments. Detailed descriptions of the quality control procedures are provided in Appendix~\ref{Details of Quality Control}.

\section{Experiment Settings}
\label{sec4}
\subsection{Baseline Models}
We tested 10 mainstream MLLMs. Closed-source models were accessed via their respective APIs, while open-source models were deployed locally. All inference tasks were run on 8$\times$NVIDIA A800 GPUs, using vLLM for efficient local deployment and inference (Llama used LMDeploy). The evaluation encompassed 4 closed-source models, including GPT-5.1~\cite{openai2025GPT-5.1}, Gemini-3-pro-preview ~\cite{Gemini-3-pro-preview}, Grok-4.1~\cite{xAI2025Grok-4.1}, and Claude-Sonnet-4.5~\cite{Anthropic2025Claude-Sonnet-4.5}, alongside 6 open-source models from multiple mainstream MLLMs, including Qwen3-VL-235B-A22B-Thinking, Qwen3-VL-32B-Thinking ~\cite{yang2025qwen3}, InternVL3.5-241B-A28B, InternVL3.5-30B-A3B ~\cite{wang2025internvl3}, MiniCPM-V-4.5 ~\cite{yu2025minicpm}, and Llama-3.2-11B-Vision-Instruct ~\cite{MetaAI2025Llama-3.2}. For more details on the models, please see Appendix \ref{modeloverview}.

\begin{figure}[ht] 
    \centering
    \includegraphics[width=0.7\textwidth]{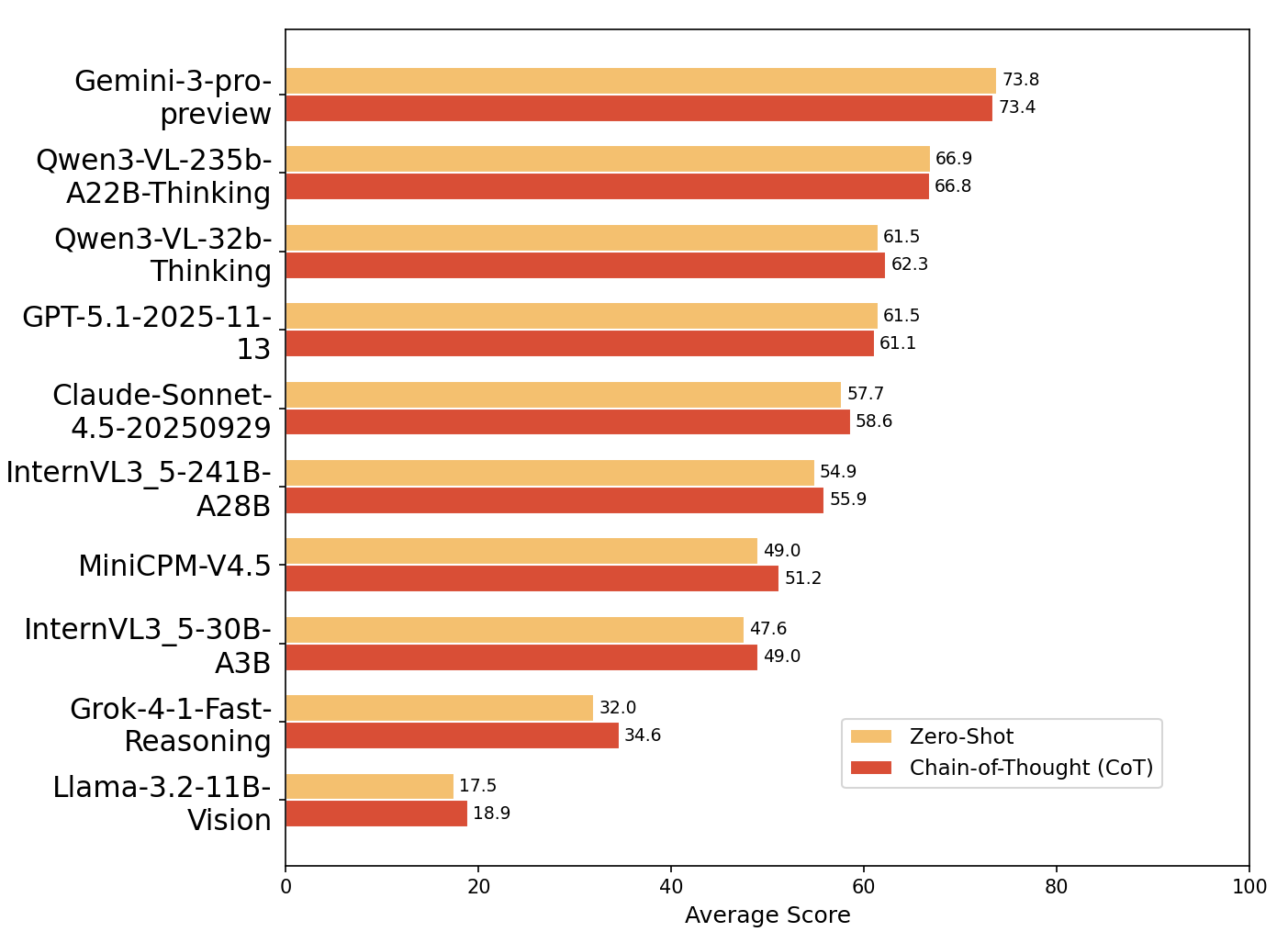} 
    \caption{As evident from the visualization of the result comparisons, the performance of the vast majority of models achieved a slight improvement under the CoT evaluation setting, though the overall magnitude of this enhancement remains relatively limited.}
    \label{fig:result}
    \vspace{-2pt}
\end{figure}

\subsection{Evaluation Methods}

We employ two core evaluation settings: Zero-Shot and Zero-Shot CoT (abbreviated as CoT). To ensure robust assessment despite the limitations of rule-based matching in complex financial reasoning, we also integrate Qwen-Max ~\cite{yang2025qwen3} to standardize output extraction and evaluation. It should be noted that a manual inspection of 30\% of the judging results from Qwen-Max revealed an average error rate of less than 1\%. Additionally, Accuracy is adopted as the core metric to ensure an objective and scalable evaluation.

\section{Results}
\label{sec5}
\subsection{Main Result}

\begin{table*}[t]
\centering
\caption{Performance of in Zero-shot and CoT settings on UniFinEval(\%). 
FSA stands for Financial Statement Auditing, CFR denotes Company Fundamental Reasoning, 
ITI refers to Industry Trend Insights, FRS represents Financial Risk Sensing, 
and AAA corresponds to Asset Allocation Analysis.The highest values in each column are highlighted with a \colorbox[HTML]{d6e7f5}{\textbf{blue background}}, while the second-best results are \underline{underlined}.}
\label{tab:cot result}

\resizebox{\textwidth}{!}{
\begin{tabular}{l c c c c c c c c c c c c}
\toprule
\multirow{2}{*}{\textbf{Model}} &
\multicolumn{2}{c}{\textbf{FSA}} & 
\multicolumn{2}{c}{\textbf{CFR}} &
\multicolumn{2}{c}{\textbf{ITI}} &
\multicolumn{2}{c}{\textbf{FRS}} &
\multicolumn{2}{c}{\textbf{AAA}} &
\multicolumn{2}{c}{\textbf{Average}}\\
\cmidrule(lr){2-3}\cmidrule(lr){4-5}\cmidrule(lr){6-7}
\cmidrule(lr){8-9}\cmidrule(lr){10-11}\cmidrule(lr){12-13}
& Zero-Shot & CoT & Zero-Shot & CoT & Zero-Shot & CoT & Zero-Shot & CoT & Zero-Shot & CoT & Zero-Shot & CoT\\
\midrule

Gemini-3-pro-preview
& \cellcolor[HTML]{d6e7f5}\textbf{83.5} & \cellcolor[HTML]{d6e7f5}\textbf{83.8}
& \cellcolor[HTML]{d6e7f5}\textbf{82.2} & \cellcolor[HTML]{d6e7f5}\textbf{82.8}
& \cellcolor[HTML]{d6e7f5}\textbf{73.3} & \cellcolor[HTML]{d6e7f5}\textbf{74.7}
& \cellcolor[HTML]{d6e7f5}\textbf{68.8} & \cellcolor[HTML]{d6e7f5}\textbf{70.1}
& \cellcolor[HTML]{d6e7f5}\textbf{61.1} & \cellcolor[HTML]{d6e7f5}\textbf{55.4}
& \cellcolor[HTML]{d6e7f5}\textbf{73.8} & \cellcolor[HTML]{d6e7f5}\textbf{73.4} \\

Qwen3-VL-235B-A22B-Thinking
& \underline{80.2} & \underline{81.3}
& \underline{78.9} & \underline{74.9}
& \underline{69.4} & 64.6
& \underline{62.9} & \underline{62.7}
& 43.3 & \underline{50.3}
& \underline{66.9} & \underline{66.8} \\

Qwen3-VL-32B-Thinking
& 75.1 & 76.2
& 71.0 & 70.3
& 65.6 & \underline{65.2}
& 54.8 & 56.6
& 40.8 & 43.3
& 61.5 & 62.3 \\

GPT-5.1
& 76.9 & 77.8
& 67.1 & 65.0
& 65.8 & 60.4
& 50.0 & 54.1
& \underline{47.8} & 48.4
& 61.5 & 61.1 \\

Claude-Sonnet-4.5
& 70.8 & 71.9
& 65.4 & 68.2
& 61.7 & 61.4
& 50.0 & 50.6
& 40.8 & 42.0
& 57.7 & 58.6 \\

InternVL3.5-241B-A28B
& 69.0 & 70.6
& 66.2 & 68.7
& 63.8 & 63.8
& 37.1 & 36.2
& 38.2 & 40.1
& 54.9 & 55.9 \\

MiniCPM-V-4.5
& 65.9 & 66.2
& 62.3 & 64.1
& 53.2 & 57.9
& 30.6 & 38.0
& 33.1 & 29.9
& 49.0 & 51.2 \\

InternVL3.5-30B-A3B
& 61.5 & 61.7
& 64.7 & 59.9
& 50.0 & 52.7
& 33.9 & 35.8
& 28.0 & 34.4
& 47.6 & 49.0 \\

Grok-4.1-Fast-Reasoning
& 50.3 & 52.5
& 43.1 & 44.1
& 32.5 & 34.9
& 16.1 & 19.3
& 17.8 & 22.3
& 32.0 & 34.6 \\

Llama-3.2-11B-Vision
& 22.2 & 23.1
& 20.9 & 23.7
& 19.0 & 21.4
& 14.1 & 15.7
& 11.5 & 10.8
& 17.5 & 18.9 \\
\midrule
Expert
& \multicolumn{2}{c}{97.5}
& \multicolumn{2}{c}{95.3} 
& \multicolumn{2}{c}{90.1}
& \multicolumn{2}{c}{88.5} 
& \multicolumn{2}{c}{85.2} 
& \multicolumn{2}{c}{91.3} \\
    
\bottomrule
\end{tabular}
}
\end{table*}

Table~\ref{tab:cot result} presents the specific performance and average results of each participating model across various tasks under both Zero-Shot and CoT settings, while Figure~\ref{fig:result} provides a more intuitive visualization for comparison. Under the Zero-Shot setting, Gemini-3-pro-preview demonstrates a stable and comprehensive performance advantage across all tasks, ranking first with an average accuracy of 73.8\%. Qwen3-VL-235B-Thinking follows closely with an average accuracy of 66.9\%, a gap of only 6.9\%. This result suggests that the technical divide between current large-scale open-source models and closed-source models is progressively narrowing. Qwen3-VL-32B-Thinking and GPT-5.1 constitute the second tier; both perform robustly in perception-based tasks and certain reasoning tasks but still face significant bottlenecks in high-level decision-making. Notably, these two models achieve identical average accuracies, not only surpassing Claude-Sonnet-4.5 and InternVL3.5-241B-A28B but also demonstrating superior performance across various sub-tasks, highlighting their relatively stronger comprehensive competitiveness. In contrast, the remaining models exhibit a clear performance gap, a stratification effect that is further amplified across different financial tasks. Model performance shows a significant gradient decay as task cognitive complexity increases. In perception-oriented tasks such as FSA and CFR, the performance gap between top-tier models and human experts is narrow, with models able to consistently identify most key information within complex research reports and charts. However, as tasks shift from explicit information recognition to the repeated verification of cross-modal information, all models experience a marked decline in performance. Even the top-performing Gemini-3-pro-preview achieves only 73.3\% accuracy in ITI scenarios, significantly lower than the 90.1\% achieved by human experts, exposing deficiencies in building consistent semantic mappings between different financial modalities. In the FRS task, which incorporate video modalities, most models fail to achieve a performance breakthrough, indicating a clear weakness in their ability to model logic across the temporal dimension.

When task complexity further escalates to AAA tasks, the performance shortfalls of all models are fully exposed: Gemini-3-pro-preview achieves an accuracy of only 61.1\%, while the performance of other models is even more inferior, creating a significant gap compared to the 85.2\% accuracy of human experts. This result clearly illustrates that although some models demonstrate strong capabilities in preliminary perception and reasoning tasks, they struggle to maintain long-term, stable logical consistency in complex financial multimodal scenarios with high information density. In contrast, human experts maintain a significant advantage across all task scenarios, and this advantage continues to expand as tasks progress from the perception level to the decision-making level. This profoundly reveals the critical capability gap that exists between current MLLMs and real-world financial experts.

From the perspective of comparison with the CoT setting, most models achieve a certain degree of performance improvement, though the magnitude of this improvement remains relatively limited. This phenomenon suggests that the vast majority of current MLLMs are already capable of adapting to explicit reasoning mechanisms; even without embedded explicit reasoning instructions, the comprehensive performance of these models can be relatively fully unleashed.

\subsection{Error Analysis}

We conducted a systematic error analysis by randomly sampling approximately 50\% of all incorrect predictions generated by the evaluated MLLMs. The overall error types are categorized into five core dimensions: Perception, Hallucination, Knowledge, Cross-modal, and Computation. These correspond to specific errors including Financial Image Perception and Data Interpretation (FIPDI), Inconsistent Financial Reasoning and Hallucinations (IFRH), Financial Knowledge Reasoning and Domain-Specific Understanding (FKRDU), Cross-modal Data Integration and Alignment (CDIA), and Financial Computation and Numerical Analysis (FCNA).

\begin{figure}[htbp]
    \centering
    \includegraphics[width=0.8\textwidth]{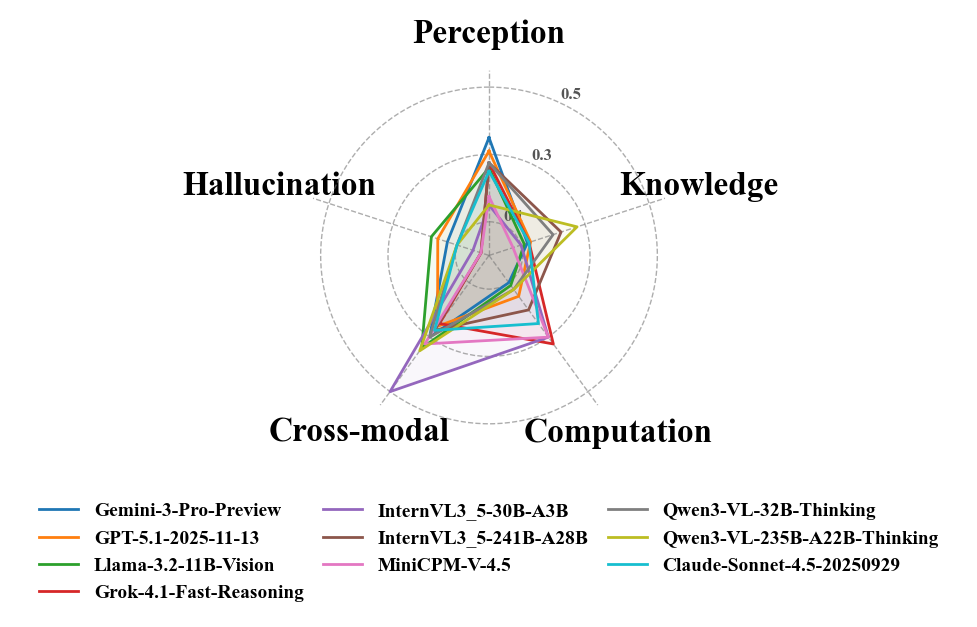}
    \caption{The radar chart summarizes the relative proportions of five major categories of errors observed in incorrect model predictions. 
    Each axis reflects the proportion of a specific error type among all erroneous cases for a given model, highlighting differences in error concentration and reasoning weaknesses across models.}
    \label{fig:financial_radar}
\end{figure}

As illustrated in Figure~\ref{fig:financial_radar}, several models still exhibit significant deficiencies in numerical computation; specifically, Qwen3-VL-32B-Thinking shows a markedly higher proportion of errors in computational tasks compared to other evaluated models. Aside from the prominent hallucination issues observed in Llama-3.2-11B-Vision and the substantial share of financial knowledge reasoning errors in Qwen3-235B-VL-A22B-Thinking, the errors of the vast majority of models are concentrated in two dimensions: image content perception and cross-modal data alignment. This phenomenon clearly demonstrates that current MLLMs still possess obvious capability shortfalls when confronting the high information density and time-sensitive complexities of financial market environments. They remain unable to fully adapt to multi-hop reasoning tasks with fine-grained requirements, failing to satisfy the rigorous demands for precision and stability inherent in the financial domain. Given that the overall performance of Grok-4.1 fell short of expectations, we conducted a targeted error analysis in Figure~\ref{fig:grok1} in Appendix~\ref{error analysis} to substantiate the finding that some general-purpose MLLMs exhibit weak adaptability in the financial domain. Simultaneously, Appendix~\ref{error analysis} provides supplementary case studies of representative error types to further support our research conclusions.

\section{Conclusion}
\label{sec6}
We proposed UniFinEval, a high-information-density multimodal benchmark oriented towards real financial business scenarios, aimed at systematically characterizing the capability boundaries of MLLMs in financial scenarios. UniFinEval centers on five scenarios: Financial Statement Auditing, Company Fundamental Reasoning, Industry Trend Insights, Financial Risk Sensing, and Asset Allocation Analysis, covering the complete financial analysis link from basic perception to expert-level decision support. The benchmark was constructed entirely manually by financial domain experts; all questions align strictly with real business logic and explicitly introduce cross-modal consistency judgment and multi-hop information integration, thereby achieving performance evaluation of actual business capabilities. Experimental results show that current MLLMs possess strong capabilities in information extraction and fact recognition tasks, but their performance remains significantly limited and lags notably behind financial experts when performing cross-modal semantic alignment, subjective judgment understanding, and complex financial decision-making under high-information-density conditions. Error analysis further revealed six major shortcomings of MLLMs in real financial scenarios.Overall, UniFinEval contributes a unified and practical evaluation framework for systematically assessing the real-world effectiveness of MLLMs in financial scenarios.

\section*{Limitations}
Although UniFinEval strives to align with real financial business processes in scenario design and capability characterization, and has made breakthrough progress in evaluating MLLMs in the financial domain, some limitations remain. First although the benchmark systematically introduces cross-modal consistency, multi-hop reasoning, and multi-turn information integration requirements in various scenarios, current tasks are still dominated by offline Q\&A forms, not yet fully simulating the complex closed loop of long-duration interaction and dynamic decision feedback in real analysis processes. Second, evaluation results mainly focus on the correctness and consistency of model outputs, leaving room for deepening the analysis of the interpretability of model reasoning paths.

\section*{Acknowledgments}
This work was supported by the National Social Science Fund of China Project under Grant No. 22BTJ031; and the Shanghai Engineering Research Center of Finance Intelligence under Grant No. 19DZ2254600. We acknowledge the technical support from the Qinghai Provincial Key Laboratory of Big Data in Finance and Artificial Intelligence Application Technology.



\bibliography{custom}

@inproceedings{tang2025finmmr,
  title={Finmmr: make financial numerical reasoning more multimodal, comprehensive, and challenging},
  author={Tang, Zichen and Liu, Jiacheng and Yang, Zhongjun and Li, Rongjin and Rong, Zihua and He, Haoyang and Hao, Zhuodi and Hu, Xinyang and Ji, Kun and Ma, Ziyan and others},
  booktitle={Proceedings of the IEEE/CVF International Conference on Computer Vision},
  pages={3245--3257},
  year={2025}
}

@article{rahman2025llm,
  title={Llm-based data science agents: A survey of capabilities, challenges, and future directions},
  author={Rahman, Mizanur and Bhuiyan, Amran and Islam, Mohammed Saidul and Laskar, Md Tahmid Rahman and Mahbub, Ridwan and Masry, Ahmed and Joty, Shafiq and Hoque, Enamul},
  journal={arXiv preprint arXiv:2510.04023},
  year={2025}
}

@article{li2025survey,
  title={A Survey on Open Dataset Search in the LLM Era: Retrospectives and Perspectives},
  author={Li, Pengyue and Wang, Sheng and Dai, Hua and Chen, Zhiyu and Bao, Zhifeng and Davison, Brian D},
  journal={arXiv preprint arXiv:2509.00728},
  year={2025}
}

@inproceedings{zhang2025data,
  title={Data cleaning using large language models},
  author={Zhang, Shuo and Huang, Zezhou and Wu, Eugene},
  booktitle={2025 IEEE 41st International Conference on Data Engineering Workshops (ICDEW)},
  pages={28--32},
  year={2025},
  organization={IEEE}
}

@article{dennstadt2025comparative,
  title={A Comparative Performance Analysis of Regular Expressions and an LLM-Based Approach to Extract the BI-RADS Score from Radiological Reports},
  author={Dennst{\"a}dt, Fabio and Lerch, Luc and Schmerder, Max and Cihoric, Nikola and Cereghetti, Grazia Maria and Gaio, Roberto and Bonel, Harald and Filchenko, Irina and Hastings, Janna and Dammann, Florian and others},
  journal={medRxiv},
  pages={2025--06},
  year={2025},
  publisher={Cold Spring Harbor Laboratory Press}
}

@article{xia2024sportu,
  title={Sportu: A comprehensive sports understanding benchmark for multimodal large language models},
  author={Xia, Haotian and Yang, Zhengbang and Zou, Junbo and Tracy, Rhys and Wang, Yuqing and Lu, Chi and Lai, Christopher and He, Yanjun and Shao, Xun and Xie, Zhuoqing and others},
  journal={arXiv preprint arXiv:2410.08474},
  year={2024}
}

@article{luo2025finmme,
  title={FinMME: Benchmark Dataset for Financial Multi-Modal Reasoning Evaluation},
  author={Luo, Junyu and Kou, Zhizhuo and Yang, Liming and Luo, Xiao and Huang, Jinsheng and Xiao, Zhiping and Peng, Jingshu and Liu, Chengzhong and Ji, Jiaming and Liu, Xuanzhe and others},
  journal={arXiv preprint arXiv:2505.24714},
  year={2025}
}

@inproceedings{deng2025finmr,
  title={FinMR: A Knowledge-Intensive Multimodal Benchmark for Advanced Financial Reasoning},
  author={Deng, Shuangyan and Peng, Haizhou and Xu, Jiachen and Mao, Rui and Giurcaneanu, Ciprian Doru and Liu, Jiamou},
  booktitle={Proceedings of the 6th ACM International Conference on AI in Finance},
  pages={168--176},
  year={2025}
}

@article{shu2025finchart,
  title={Finchart-bench: Benchmarking financial chart comprehension in vision-language models},
  author={Shu, Dong and Yuan, Haoyang and Wang, Yuchen and Liu, Yanguang and Zhang, Huopu and Zhao, Haiyan and Du, Mengnan},
  journal={arXiv preprint arXiv:2507.14823},
  year={2025}
}

@article{peng2025multifinben,
  title={MultiFinBen: A Multilingual, Multimodal, and Difficulty-Aware Benchmark for Financial LLM Evaluation},
  author={Peng, Xueqing and Qian, Lingfei and Wang, Yan and Xiang, Ruoyu and He, Yueru and Ren, Yang and Jiang, Mingyang and Zhao, Jeff and He, Huan and Han, Yi and others},
  journal={arXiv preprint arXiv:2506.14028},
  year={2025}
}

@article{li2025cfbenchmark,
  title={CFBenchmark-MM: Chinese Financial Assistant Benchmark for Multimodal Large Language Model},
  author={Li, Jiangtong and Zhu, Yiyun and Cheng, Dawei and Ding, Zhijun and Jiang, Changjun},
  journal={arXiv preprint arXiv:2506.13055},
  year={2025}
}

@inproceedings{zhang2025xfinbench,
  title={XFINBENCH: Benchmarking llms in complex financial problem solving and reasoning},
  author={Zhang, Zhihan and Cao, Yixin and Liao, Lizi},
  booktitle={Findings of the Association for Computational Linguistics: ACL 2025},
  pages={8715--8758},
  year={2025}
}

@article{xu2025finmultitime,
  title={Finmultitime: A four-modal bilingual dataset for financial time-series analysis},
  author={Xu, Wenyan and Xiang, Dawei and Liu, Yue and Wang, Xiyu and Ma, Yanxiang and Zhang, Liang and Hu, Shu and Xu, Chang and Zhang, Jiaheng},
  journal={arXiv preprint arXiv:2506.05019},
  year={2025}
}

@inproceedings{galarnyk2025videoconviction,
  title={Videoconviction: A multimodal benchmark for human conviction and stock market recommendations},
  author={Galarnyk, Michael and Kejriwal, Veer and Shah, Agam and Bhardwaj, Yash and Meyer, Nicholas Watney and Krishnan, Anand and Chava, Sudheer},
  booktitle={Proceedings of the 31st ACM SIGKDD Conference on Knowledge Discovery and Data Mining V. 2},
  pages={5447--5458},
  year={2025}
}

@article{chen2025mtbench,
  title={Mtbench: A multimodal time series benchmark for temporal reasoning and question answering},
  author={Chen, Jialin and Feng, Aosong and Zhao, Ziyu and Garza, Juan and Nurbek, Gaukhar and Qin, Cheng and Maatouk, Ali and Tassiulas, Leandros and Gao, Yifeng and Ying, Rex},
  journal={arXiv preprint arXiv:2503.16858},
  year={2025}
}

@article{li2024seed,
  title={Seed-bench-2-plus: Benchmarking multimodal large language models with text-rich visual comprehension},
  author={Li, Bohao and Ge, Yuying and Chen, Yi and Ge, Yixiao and Zhang, Ruimao and Shan, Ying},
  journal={arXiv preprint arXiv:2404.16790},
  year={2024}
}

@inproceedings{liu2025findabench,
  title={Findabench: Benchmarking financial data analysis ability of large language models},
  author={Liu, Shu and Zhao, Shangqing and Jia, Chenghao and Zhuang, Xinlin and Long, Zhaoguang and Zhou, Jie and Zhou, Aimin and Lan, Man and Chong, Yang},
  booktitle={Proceedings of the 31st International Conference on Computational Linguistics},
  pages={710--725},
  year={2025}
}

@article{zhang2024mme,
  title={Mme-realworld: Could your multimodal llm challenge high-resolution real-world scenarios that are difficult for humans?},
  author={Zhang, Yi-Fan and Zhang, Huanyu and Tian, Haochen and Fu, Chaoyou and Zhang, Shuangqing and Wu, Junfei and Li, Feng and Wang, Kun and Wen, Qingsong and Zhang, Zhang and others},
  journal={arXiv preprint arXiv:2408.13257},
  year={2024}
}

@article{chen2025mm,
  title={MM-DREX: Multimodal-Driven Dynamic Routing of LLM Experts for Financial Trading},
  author={Chen, Yang and Jiang, Yueheng and Ma, Zhaozhao and Cao, Yuchen and Keung, Jacky and Kuang, Kun and Gan, Leilei and Wu, Yiquan and Wu, Fei},
  journal={arXiv preprint arXiv:2509.05080},
  year={2025}
}

@article{wang2025finzero,
  title={FinZero: Launching Multi-modal Financial Time Series Forecast with Large Reasoning Model},
  author={Wang, Yanlong and Xu, Jian and Ma, Fei and Zhang, Hongkang and Yu, Hang and Gao, Tiantian and Wang, Yu and You, Haochen and Huang, Shao-Lun and Sun, Danny Dongning and others},
  journal={arXiv preprint arXiv:2509.08742},
  year={2025}
}

@inproceedings{10.1145/3746027.3754557,
author = {Das, Sarmistha and Lyngkhoi, R. E. Zera Marveen and Saha, Sriparna and Maurya, Alka},
title = {Unlocking Financial Insights: An advanced Multimodal Summarization with Multimodal Output Framework for Financial Advisory Videos},
year = {2025},
isbn = {9798400720352},
publisher = {Association for Computing Machinery},
address = {New York, NY, USA},
url = {https://doi.org/10.1145/3746027.3754557},
doi = {10.1145/3746027.3754557},
booktitle = {Proceedings of the 33rd ACM International Conference on Multimedia},
pages = {11976–11985},
numpages = {10},
location = {Dublin, Ireland},
series = {MM '25}
}

@article{lee2024survey,
  title={A survey of large language models in finance (finllms)},
  author={Lee, Jean and Stevens, Nicholas and Han, Soyeon Caren and Song, Minseok},
  journal={arXiv preprint arXiv:2402.02315},
  year={2024}
}

@inproceedings{rangapur2025fin,
  title={Fin-Fact: A Benchmark Dataset for Multimodal Financial Fact-Checking and Explanation Generation},
  author={Rangapur, Aman and Wang, Haoran and Jian, Ling and Shu, Kai},
  booktitle={Companion Proceedings of the ACM on Web Conference 2025},
  pages={785--788},
  year={2025}
}

@inproceedings{liu-etal-2025-visfineval,
    title = "{V}is{F}in{E}val: A Scenario-Driven {C}hinese Multimodal Benchmark for Holistic Financial Understanding",
    author = "Liu, Zhaowei  and
      Guo, Xin  and
      Xia, Haotian  and
      Zeng, Lingfeng  and
      Lou, Fangqi  and
      Niu, Jinyi  and
      Li, Mengping  and
      Qi, Qi  and
      Li, Jiahuan  and
      Zhang, Wei  and
      Wang, Yinglong  and
      Cai, Weige  and
      Shen, Weining  and
      Zhang, Liwen",
    editor = "Christodoulopoulos, Christos  and
      Chakraborty, Tanmoy  and
      Rose, Carolyn  and
      Peng, Violet",
    booktitle = "Proceedings of the 2025 Conference on Empirical Methods in Natural Language Processing",
    month = nov,
    year = "2025",
    address = "Suzhou, China",
    publisher = "Association for Computational Linguistics",
    url = "https://aclanthology.org/2025.emnlp-main.1229/",
    doi = "10.18653/v1/2025.emnlp-main.1229",
    pages = "24099--24157",
    ISBN = "979-8-89176-332-6"
}

@article{liu2025fin,
  title={Fin-r1: A large language model for financial reasoning through reinforcement learning},
  author={Liu, Zhaowei and Guo, Xin and Lou, Fangqi and Zeng, Lingfeng and Niu, Jinyi and Wang, Zixuan and Xu, Jiajie and Cai, Weige and Yang, Ziwei and Zhao, Xueqian and others},
  journal={arXiv preprint arXiv:2503.16252},
  year={2025}
}

@article{xiao2025trading,
  title={Trading-r1: Financial trading with llm reasoning via reinforcement learning},
  author={Xiao, Yijia and Sun, Edward and Chen, Tong and Wu, Fang and Luo, Di and Wang, Wei},
  journal={arXiv preprint arXiv:2509.11420},
  year={2025}
}

@article{kim2024financial,
  title={Financial statement analysis with large language models},
  author={Kim, Alex and Muhn, Maximilian and Nikolaev, Valeri},
  journal={arXiv preprint arXiv:2407.17866},
  year={2024}
}

@inproceedings{iacovides2024finllama,
  title={Finllama: Llm-based financial sentiment analysis for algorithmic trading},
  author={Iacovides, Giorgos and Konstantinidis, Thanos and Xu, Mingxue and Mandic, Danilo},
  booktitle={Proceedings of the 5th ACM International Conference on AI in Finance},
  pages={134--141},
  year={2024}
}

@inproceedings{guo-etal-2025-fineval,
    title = "{F}in{E}val: A {C}hinese Financial Domain Knowledge Evaluation Benchmark for Large Language Models",
    author = "Guo, Xin  and
      Xia, Haotian  and
      Liu, Zhaowei  and
      Cao, Hanyang  and
      Yang, Zhi  and
      Liu, Zhiqiang  and
      Wang, Sizhe  and
      Niu, Jinyi  and
      Wang, Chuqi  and
      Wang, Yanhui  and
      Liang, Xiaolong  and
      Huang, Xiaoming  and
      Zhu, Bing  and
      Wei, Zhongyu  and
      Chen, Yun  and
      Shen, Weining  and
      Zhang, Liwen",
    editor = "Chiruzzo, Luis  and
      Ritter, Alan  and
      Wang, Lu",
    booktitle = "Proceedings of the 2025 Conference of the Nations of the Americas Chapter of the Association for Computational Linguistics: Human Language Technologies (Volume 1: Long Papers)",
    month = apr,
    year = "2025",
    address = "Albuquerque, New Mexico",
    publisher = "Association for Computational Linguistics",
    url = "https://aclanthology.org/2025.naacl-long.318/",
    doi = "10.18653/v1/2025.naacl-long.318",
    pages = "6258--6292",
    ISBN = "979-8-89176-189-6"
}

@inproceedings{li-etal-2025-investorbench,
    title = "{INVESTORBENCH}: A Benchmark for Financial Decision-Making Tasks with {LLM}-based Agent",
    author = "Li, Haohang  and
      Cao, Yupeng  and
      Yu, Yangyang  and
      Javaji, Shashidhar Reddy  and
      Deng, Zhiyang  and
      He, Yueru  and
      Jiang, Yuechen  and
      Zhu, Zining  and
      Subbalakshmi, K.p.  and
      Huang, Jimin  and
      Qian, Lingfei  and
      Peng, Xueqing  and
      Suchow, Jordan W.  and
      Xie, Qianqian",
    editor = "Che, Wanxiang  and
      Nabende, Joyce  and
      Shutova, Ekaterina  and
      Pilehvar, Mohammad Taher",
    booktitle = "Proceedings of the 63rd Annual Meeting of the Association for Computational Linguistics (Volume 1: Long Papers)",
    month = jul,
    year = "2025",
    address = "Vienna, Austria",
    publisher = "Association for Computational Linguistics",
    url = "https://aclanthology.org/2025.acl-long.126/",
    doi = "10.18653/v1/2025.acl-long.126",
    pages = "2509--2525",
    ISBN = "979-8-89176-251-0"
}

@inproceedings{matlin-etal-2025-financial,
    title = "Financial Language Model Evaluation ({FL}a{ME})",
    author = "Matlin, Glenn  and
      Okamoto, Mika  and
      Pardawala, Huzaifa  and
      Yang, Yang  and
      Chava, Sudheer",
    editor = "Che, Wanxiang  and
      Nabende, Joyce  and
      Shutova, Ekaterina  and
      Pilehvar, Mohammad Taher",
    booktitle = "Findings of the Association for Computational Linguistics: ACL 2025",
    month = jul,
    year = "2025",
    address = "Vienna, Austria",
    publisher = "Association for Computational Linguistics",
    url = "https://aclanthology.org/2025.findings-acl.1164/",
    doi = "10.18653/v1/2025.findings-acl.1164",
    pages = "22633--22679",
    ISBN = "979-8-89176-256-5"
}

@article{yu2024mm,
  title={Mm-vet v2: A challenging benchmark to evaluate large multimodal models for integrated capabilities},
  author={Yu, Weihao and Yang, Zhengyuan and Ren, Lingfeng and Li, Linjie and Wang, Jianfeng and Lin, Kevin and Lin, Chung-Ching and Liu, Zicheng and Wang, Lijuan and Wang, Xinchao},
  journal={arXiv preprint arXiv:2408.00765},
  year={2024}
}

@article{sukhani2025fincap,
  title={FinCap: Topic-Aligned Captions for Short-Form Financial YouTube Videos},
  author={Sukhani, Siddhant and Bhardwaj, Yash and Bhadani, Riya and Kejriwal, Veer and Galarnyk, Michael and Chava, Sudheer},
  journal={arXiv preprint arXiv:2509.25745},
  year={2025}
}

@article{liu2025fingpt,
  title={AT-FinGPT: Financial risk prediction via an audio-text large language model},
  author={Liu, Yingnan and Bu, Ningbo and Li, Zhiqiang and Zhang, Yongmin and Zhao, Zhenyu},
  journal={Finance Research Letters},
  volume={77},
  pages={106967},
  year={2025},
  publisher={Elsevier}
}

@article{cao2024risklabs,
  title={RiskLabs: predicting financial risk using large language model based on multimodal and multi-sources data},
  author={Cao, Yupeng and Chen, Zhi and Kumar, Prashant and Pei, Qingyun and Yu, Yangyang and Li, Haohang and Dimino, Fabrizio and Ausiello, Lorenzo and Subbalakshmi, KP and Ndiaye, Papa Momar},
  journal={arXiv preprint arXiv:2404.07452},
  year={2024}
}

@article{masry2024longfin,
  title={LongFin: A Multimodal Document Understanding Model for Long Financial Domain Documents},
  author={Masry, Ahmed and Hajian, Amir},
  journal={arXiv preprint arXiv:2401.15050},
  year={2024}
}

@inproceedings{wilson2024fin2sum,
  title={FIN2SUM: advancing AI-driven financial text summarization with LLMs},
  author={Wilson, Ezhilan and Saxena, Anshul and Mahajan, Jayant and Panikulangara, Lekha and Kulkarni, Shruti and Jain, Pritty},
  booktitle={2024 International Conference on Trends in Quantum Computing and Emerging Business Technologies},
  pages={1--5},
  year={2024},
  organization={IEEE}
}

@article{delgadillo2024finsosent,
  title={FinSoSent: Advancing Financial Market Sentiment Analysis through Pretrained Large Language Models},
  author={Delgadillo, Josiel and Kinyua, Johnson and Mutigwe, Charles},
  journal={Big Data and Cognitive Computing},
  volume={8},
  number={8},
  pages={87},
  year={2024},
  publisher={MDPI AG}
}

@article{li2024alphafin,
  title={AlphaFin: Benchmarking Financial Analysis with Retrieval-Augmented Stock-Chain Framework},
  author={Li, Xiang and Li, Zhenyu and Shi, Chen and Xu, Yong and Du, Qing and Tan, Mingkui and Huang, Jun and Lin, Wei},
  journal={arXiv preprint arXiv:2403.12582},
  year={2024}
}

@inproceedings{li2024finreport,
  title={FinReport: Explainable Stock Earnings Forecasting via News Factor Analyzing Model},
  author={Li, Xiangyu and Shen, Xinjie and Zeng, Yawen and Xing, Xiaofen and Xu, Jin},
  booktitle={Companion Proceedings of the ACM on Web Conference 2024},
  pages={319--327},
  year={2024}
}

@article{wang2024quantagent,
  title={QuantAgent: Seeking Holy Grail in Trading by Self-Improving Large Language Model},
  author={Wang, Saizhuo and Yuan, Hang and Ni, Lionel M and Guo, Jian},
  journal={arXiv preprint arXiv:2402.03755},
  year={2024}
}

@article{mai2024stockgpt,
  title={StockGPT: A GenAI Model for Stock Prediction and Trading},
  author={Mai, Dat},
  journal={arXiv preprint arXiv:2404.05101},
  year={2024}
}

@article{yu2024fincon,
  title={Fincon: A synthesized llm multi-agent system with conceptual verbal reinforcement for enhanced financial decision making},
  author={Yu, Yangyang and Yao, Zhiyuan and Li, Haohang and Deng, Zhiyang and Cao, Yupeng and Chen, Zhi and Suchow, Jordan W and Liu, Rong and Cui, Zhenyu and Xu, Zhaozhuo and others},
  journal={arXiv preprint arXiv:2407.06567},
  year={2024}
}

@inproceedings{yu2024finmem,
  title={FinMem: A performance-enhanced LLM trading agent with layered memory and character design},
  author={Yu, Yangyang and Li, Haohang and Chen, Zhi and Jiang, Yuechen and Li, Yang and Zhang, Denghui and Liu, Rong and Suchow, Jordan W and Khashanah, Khaldoun},
  booktitle={Proceedings of the AAAI Symposium Series},
  volume={3},
  pages={595--597},
  year={2024}
}

@article{nie2024cfinbench,
  title={CFinBench: A Comprehensive Chinese Financial Benchmark for Large Language Models},
  author={Nie, Ying and Yan, Binwei and Guo, Tianyu and Liu, Hao and Wang, Haoyu and He, Wei and Zheng, Binfan and Wang, Weihao and Li, Qiang and Sun, Weijian and others},
  journal={arXiv preprint arXiv:2407.02301},
  year={2024}
}

@article{chen2021finqa,
  title={Finqa: A dataset of numerical reasoning over financial data},
  author={Chen, Zhiyu and Chen, Wenhu and Smiley, Charese and Shah, Sameena and Borova, Iana and Langdon, Dylan and Moussa, Reema and Beane, Matt and Huang, Ting-Hao and Routledge, Bryan and others},
  journal={arXiv preprint arXiv:2109.00122},
  year={2021}
}

@article{zhu2024benchmarking,
  title={Benchmarking Large Language Models on CFLUE--A Chinese Financial Language Understanding Evaluation Dataset},
  author={Zhu, Jie and Li, Junhui and Wen, Yalong and Guo, Lifan},
  journal={arXiv preprint arXiv:2405.10542},
  year={2024}
}

@article{wang2023finvis,
  title={FinVis-GPT: A multimodal large language model for financial chart analysis},
  author={Wang, Ziao and Li, Yuhang and Wu, Junda and Soon, Jaehyeon and Zhang, Xiaofeng},
  journal={arXiv preprint arXiv:2308.01430},
  year={2023}
}

@article{zhang2024finagent,
  title={FinAgent: A Multimodal Foundation Agent for Financial Trading: Tool-Augmented, Diversified, and Generalist},
  author={Zhang, Wentao and Zhao, Lingxuan and Xia, Haochong and Sun, Shuo and Sun, Jiaze and Qin, Molei and Li, Xinyi and Zhao, Yuqing and Zhao, Yilei and Cai, Xinyu and others},
  journal={arXiv preprint arXiv:2402.18485},
  year={2024}
}

@article{bhatia2024fintral,
  title={Fintral: A family of gpt-4 level multimodal financial large language models},
  author={Bhatia, Gagan and Nagoudi, El Moatez Billah and Cavusoglu, Hasan and Abdul-Mageed, Muhammad},
  journal={arXiv preprint arXiv:2402.10986},
  year={2024}
}

@inproceedings{zhao2024financemath,
  title={Financemath: Knowledge-intensive math reasoning in finance domains},
  author={Zhao, Yilun and Liu, Hongjun and Long, Yitao and Zhang, Rui and Zhao, Chen and Cohan, Arman},
  booktitle={Proceedings of the 62nd Annual Meeting of the Association for Computational Linguistics (Volume 1: Long Papers)},
  pages={12841--12858},
  year={2024}
}

@inproceedings{wang2024doctabqa,
  title={DocTabQA: Answering Questions from Long Documents Using Tables},
  author={Wang, Haochen and Hu, Kai and Dong, Haoyu and Gao, Liangcai},
  booktitle={International Conference on Document Analysis and Recognition},
  pages={470--487},
  year={2024},
  organization={Springer}
}

@article{reddy2024docfinqa,
  title={Docfinqa: A long-context financial reasoning dataset},
  author={Reddy, Varshini and Koncel-Kedziorski, Rik and Lai, Viet Dac and Krumdick, Michael and Lovering, Charles and Tanner, Chris},
  journal={arXiv preprint arXiv:2401.06915},
  year={2024}
}

@article{chen2024fintextqa,
  title={FinTextQA: A Dataset for Long-form Financial Question Answering},
  author={Chen, Jian and Zhou, Peilin and Hua, Yining and Loh, Yingxin and Chen, Kehui and Li, Ziyuan and Zhu, Bing and Liang, Junwei},
  journal={arXiv preprint arXiv:2405.09980},
  year={2024}
}

@inproceedings{liu2024mmbench,
  title={Mmbench: Is your multi-modal model an all-around player?},
  author={Liu, Yuan and Duan, Haodong and Zhang, Yuanhan and Li, Bo and Zhang, Songyang and Zhao, Wangbo and Yuan, Yike and Wang, Jiaqi and He, Conghui and Liu, Ziwei and others},
  booktitle={European conference on computer vision},
  pages={216--233},
  year={2024},
  organization={Springer}
}

@inproceedings{yue2024mmmu,
  title={Mmmu: A massive multi-discipline multimodal understanding and reasoning benchmark for expert agi},
  author={Yue, Xiang and Ni, Yuansheng and Zhang, Kai and Zheng, Tianyu and Liu, Ruoqi and Zhang, Ge and Stevens, Samuel and Jiang, Dongfu and Ren, Weiming and Sun, Yuxuan and others},
  booktitle={Proceedings of the IEEE/CVF Conference on Computer Vision and Pattern Recognition},
  pages={9556--9567},
  year={2024}
}

@article{xue2024famma,
  title={FAMMA: A Benchmark for Financial Domain Multilingual Multimodal Question Answering},
  author={Xue, Siqiao and Chen, Tingting and Zhou, Fan and Dai, Qingyang and Chu, Zhixuan and Mei, Hongyuan},
  journal={arXiv preprint arXiv:2410.04526},
  year={2024}
}

@article{gan2024mme,
  title={MME-Finance: A Multimodal Finance Benchmark for Expert-level Understanding and Reasoning},
  author={Gan, Ziliang and Lu, Yu and Zhang, Dong and Li, Haohan and Liu, Che and Liu, Jian and Liu, Ji and Wu, Haipang and Fu, Chaoyou and Xu, Zenglin and others},
  journal={arXiv preprint arXiv:2411.03314},
  year={2024}
}

@misc{openai2025GPT-5.1,
  title={GPT-5.1: A smarter, more conversational ChatGPT},
  author={OpenAI},
  howpublished={\url{https://openai.com/zh-Hans-CN/index/gpt-5-1/}},
  year={2025}
}

@misc{Gemini-3-pro-preview,
  title={Gemini-3-pro-preview},
  author={Google},
  year         = {2025},
  publisher    = {Google},
  howpublished = {\url{https://gemini.google.com/}},
}

@misc{xAI2025Grok-4.1,
  title={Grok 4.1 Fast and Agent Tools API},
  author={xAI},
  howpublished={\url{https://x.ai/grok}},
  year={2025}
}

@misc{Anthropic2025Claude-Sonnet-4.5,
  title={Claude-Sonnet-4.5},
  author={Anthropic},
  howpublished={\url{https://www.anthropic.com/claude/sonnet}},
  year={2025}
}

@article{yang2025qwen3,
  title={Qwen3 technical report},
  author={Yang, An and Li, Anfeng and Yang, Baosong and Zhang, Beichen and Hui, Binyuan and Zheng, Bo and Yu, Bowen and Gao, Chang and Huang, Chengen and Lv, Chenxu and others},
  journal={arXiv preprint arXiv:2505.09388},
  year={2025}
}

@article{wang2025internvl3,
  title={Internvl3. 5: Advancing open-source multimodal models in versatility, reasoning, and efficiency},
  author={Wang, Weiyun and Gao, Zhangwei and Gu, Lixin and Pu, Hengjun and Cui, Long and Wei, Xingguang and Liu, Zhaoyang and Jing, Linglin and Ye, Shenglong and Shao, Jie and others},
  journal={arXiv preprint arXiv:2508.18265},
  year={2025}
}

@misc{MetaAI2025Llama-3.2,
  title={Llama 3.2: Revolutionizing edge AI and vision with open, customizable models},
  author={MetaAI},
  howpublished={\url{https://huggingface.co/meta-llama/Llama-3.2-11B-Vision-Instruct}},
  year={2025}
}

@article{yu2025minicpm,
  title={Minicpm-v 4.5: Cooking efficient mllms via architecture, data, and training recipe},
  author={Yu, Tianyu and Wang, Zefan and Wang, Chongyi and Huang, Fuwei and Ma, Wenshuo and He, Zhihui and Cai, Tianchi and Chen, Weize and Huang, Yuxiang and Zhao, Yuanqian and others},
  journal={arXiv preprint arXiv:2509.18154},
  year={2025}
}

\newpage
\clearpage
\appendix

\clearpage

\centerline{\maketitle{\textbf{SUMMARY OF THE APPENDIX}}}

This appendix contains additional details for the  \textbf{\textit{``UniFinEval: Towards Unified Evaluation of Financial Multimodal Models
across Text, Images and Videos''}}. The appendix
is organized as follows:

\startcontents[appendices]
\section*{Appendix Table of Contents}
\printcontents[appendices]{}{0}{\large}

\section{Details of UniFinEval} 
\label{benchmark}

\subsection{Comparison of benchmarks}
In this section, we first provide a detailed comparison Table ~\ref{tab:benchmarks-comparison} of various Q\&A datasets across multiple dimensions. 

\begin{table*}[!htbp]
\caption{Comparison of various benchmarks across multiple dimensions. The abbreviations in the header are: MLD (Multi-level Difficulty), RES (Realistic Environment Simulation), MC (Manually Construct), FAV (Financial Analysis Video), CD (Consistency Detection), and MHR (Multi-Hop Reasoning).}
\label{tab:benchmarks-comparison}
\renewcommand{\arraystretch}{0.3} 
\setlength{\tabcolsep}{14pt} 

\begin{tabular}{
  l 
  c 
  c 
  c 
  c 
  c 
  c 
}
\toprule[2pt]
\textbf{Benchmarks} & \textbf{MLD} & \textbf{RES} & \textbf{MC} & \textbf{FAV} & \textbf{CD} & \textbf{MHR} \\
\midrule
\multicolumn{7}{c}{\textbf{Image}} \\
\midrule
FinMR & \cmark & \xmark & \xmark & \xmark & \xmark & \xmark \\
\addlinespace
FinMMR & \cmark & \xmark & \xmark & \xmark & \xmark & \xmark \\
\addlinespace
FinMME & \cmark & \xmark & \xmark & \xmark & \xmark & \xmark \\
\addlinespace
Multifinben & \cmark & \xmark & \xmark & \xmark & \xmark & \xmark \\
\addlinespace
XFinBench & \cmark & \xmark & \xmark & \xmark & \xmark & \xmark \\
\addlinespace
VisFinEval & \cmark & \cmark & \xmark & \xmark & \xmark & \xmark \\
\addlinespace
FinMultiTime & \cmark & \xmark & \xmark & \xmark & \xmark & \xmark \\
\addlinespace
CFBenchmark-MM & \xmark & \xmark & \xmark & \xmark & \xmark & \xmark \\
\addlinespace
\midrule
\multicolumn{7}{c}{\textbf{Video}} \\
\midrule
FinCap & \cmark & \xmark & \xmark & \cmark & \xmark & \xmark \\
\addlinespace
VideoConviction & \cmark & \xmark & \xmark & \cmark & \xmark & \xmark \\
\addlinespace
\midrule
\textbf{Ours(Text + Image + Video)} & \cmark & \cmark & \cmark & \cmark & \cmark & \cmark \\
\bottomrule[2pt]
\end{tabular}%
\end{table*}

\subsection{Statistic and Examples of Financial Business Scenarios}
We list the detailed information of UniFinEval data in Table~\ref{tab:Unifineval_dist}. The detailed information of the financial business cenarios are presented below.

\begin{table}[!htbp]
    \centering
    \caption{Financial Scenario Data Distribution. The table details the distribution of questions across five core financial tasks: Financial Statement Auditing
    (FSA), Company Fundamental Reasoning (CFR), Industry Trend Insights (ITI), Financial Risk Sensing (FRS), and Asset Allocation Analysis (AAA).}
    \label{tab:Unifineval_dist}
    \vspace{5pt}
        \begin{tabular}{lr}
            \toprule[2pt]
            \textbf{Financial Scenario} & \textbf{Questions} \\
            \midrule
            Financial Statement Auditing & 892 \\
            Company Fundamental Reasoning & 926 \\
            Industry Trend Insights & 896 \\
            Financial Risk Sensing & 535 \\
            Asset Allocation Analysis & 518 \\
            \midrule[1pt]
            \textbf{Total (UniFinEval)} & \textbf{3767} \\
            \bottomrule[2pt]
        \end{tabular}%
\end{table}

\begin{figure*}[htbp] 
\begin{tcolorbox}[
    colback=darkgray!5!white,
    title=Financial Statement Auditing, 
    colframe=darkgray!2!darkgray,
    arc=2mm,
    fontupper=\small,
    breakable=false,
    width=\textwidth 
]

\begin{center} 
    \includegraphics[width=0.9\linewidth]{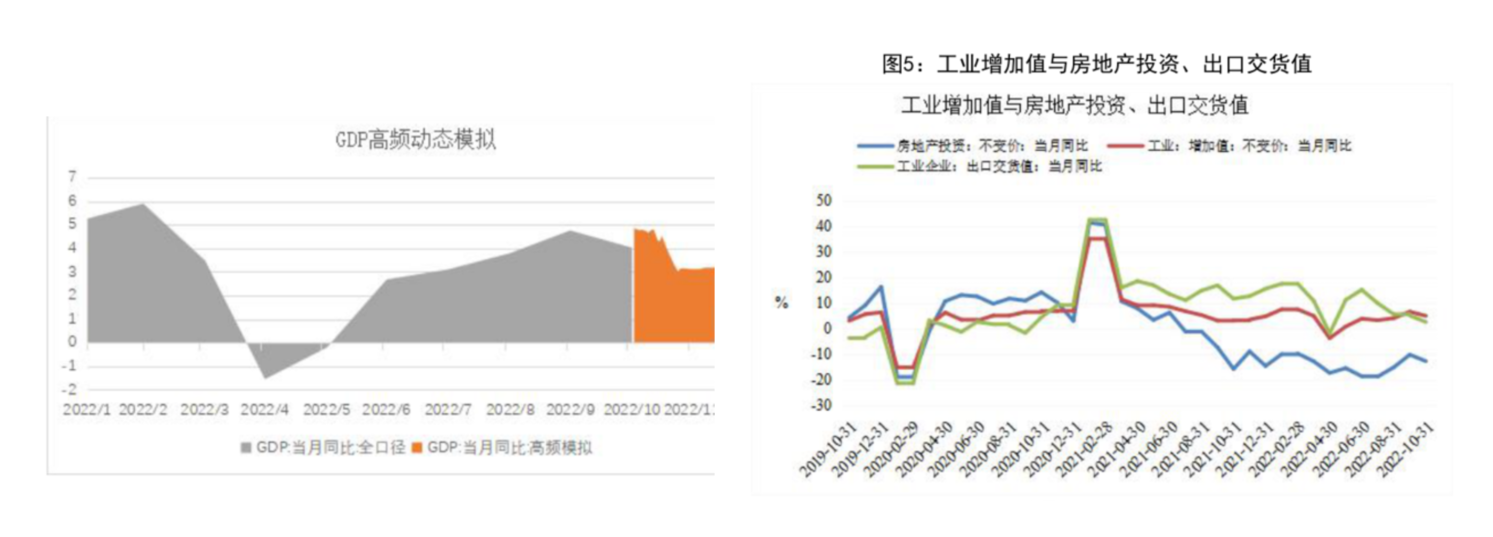}
\end{center}
\begin{CJK*}{UTF8}{gbsn}
文本：工业生产稳中向好，计算机通信动能较强腾景全口径数据显示，工业经济运行相对平稳，1-10月工业增加值不变价累计同比增速为3.8\%,其中4月份受到疫情冲击影响，增速落入负向区间(-3.9\%),而后触底反弹、保持微幅上行态势。其中，房地产投资、出口交货值与工业增加值同比增速走势契合度较高，但在2021年5月出现K形分化走势，今年K形分化程度进一步加深，具体表现在房地产投资增速自去年7月份落入负向区间后，不断向下走阔， 持续两位数的负增长低迷状态，阻碍工业修复。
\end{CJK*}

{\color{blue}
Text: Industrial production remains stable with a positive outlook, driven by strong momentum in the computer and communication sectors. According to Tengjing’s comprehensive data, the industrial economy is operating relatively steadily. From January to October, the cumulative year-on-year growth rate of industrial value-added (at constant prices) was 3.8\%. Notably, growth fell into negative territory (-3.9\%) in April due to the impact of the pandemic, but subsequently bottomed out and rebounded, maintaining a slight upward trajectory. Historically, the year-on-year growth trends of real estate investment and export delivery value have closely aligned with industrial value-added. However, a "K-shaped" divergence emerged in May 2021 and has intensified this year. This is specifically manifested in real estate investment growth: since falling into negative territory last July, the decline has widened, persisting in a slump of double-digit negative growth that is hindering industrial recovery.
}

Round 1:
\\
\begin{CJK*}{UTF8}{gbsn}
问题：依据文本中关于房地产投资增速的描述，其进入持续两位数负增长区间的起始月份是哪个月？
\end{CJK*}
{\color{blue}
Question: Based on the description of the real estate investment growth rate in the text, what is the starting month in which it entered the range of sustained double-digit negative growth?
}
\begin{CJK*}{UTF8}{gbsn}
答案：2021年7月
\end{CJK*}
{\color{blue}
Answer: July 2021
}
\\
Round 2:
\\
\begin{CJK*}{UTF8}{gbsn}
问题：结合确定的起始月份和图5中显示的最新数据点，计算房地产投资增速处于两位数负增长状态的总月份数。
\end{CJK*}

{\color{blue}
Question: Combining the identified starting month with the latest data point shown in Figure 5, calculate the total number of months during which the real estate investment growth rate remained in a state of double-digit negative growth.
}

\begin{CJK*}{UTF8}{gbsn}
答案：16
\end{CJK*}

{\color{blue}
Answer: 16
}
\\
\end{tcolorbox}
\caption{This example showcases the Financial Statement Auditing scenario, focusing on high-precision data retrieval and alignment. To provide an accurate answer, the model must demonstrate the ability to precisely localize specific indicators across unstructured text and complex time-series charts. Specifically, the model is required to pinpoint the exact starting month (July 2021) of sustained double-digit negative growth in real estate investment within the text, and then accurately map its location to the corresponding trend in Figure 5. This task evaluates the model's ability to maintain high-precisionn and consistency verification when tracking specific industry fluctuations within high-density and specialized financial materials. }
\label{fig:secnario-1}
\end{figure*}

\textbf{Financial Statement Auditing:} As the foundational entry point of financial operations, its core objective is to accurately localize key financial indicators and event-related information from high-density materials such as financial statements and research reports. This establishes a solid data foundation for all subsequent analyses and decision-making, ensuring the accuracy of information at its source.
Unlike data using only simplified charts, this scenario retains real typesetting and redundant information, forcing the model to complete key information screening under real visual interference. Question design transitions from single factn to cross-page, multi-point multi-hop reasoning, simulating the actual work process of analysts integrating information across multiple pages. This assesses whether the model possesses the ability to effectively screen and precisely localize information within real financial materials. Figure \ref{fig:secnario-1} provides examples of Financial Statement Auditing.

\begin{figure*}[htbp] 
\begin{tcolorbox}[
    colback=darkgray!5!white,
    title=Company Fundamental Reasoning, 
    colframe=darkgray!2!darkgray,
    arc=2mm,
    fontupper=\small,
    breakable=false,
    width=\textwidth 
]

\begin{center} 
    \includegraphics[width=0.7\linewidth]{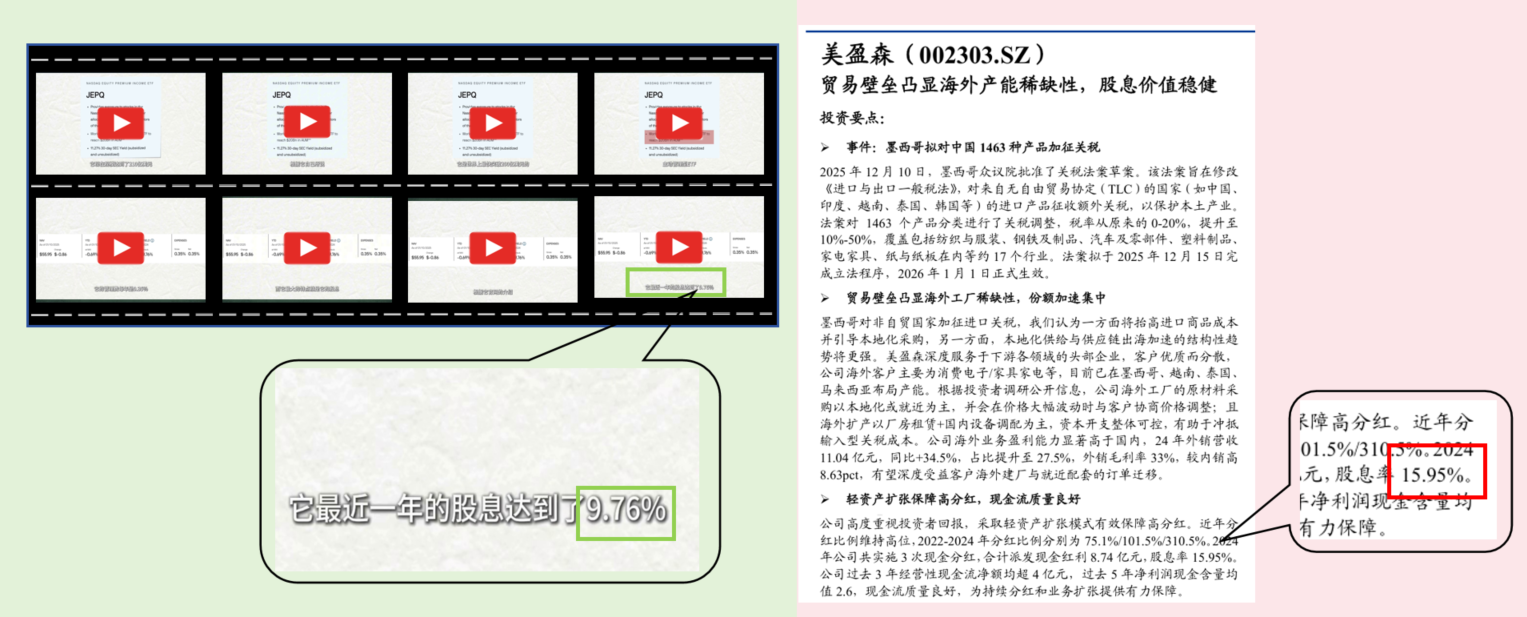}
\end{center}

\begin{CJK*}{UTF8}{gbsn}
问题：世界上规模最快突破两百亿美元的主动管理型ETF的股息与美盈森的差值是多少？
\end{CJK*}

{\color{blue}
Question: What is the difference between the dividend of the world's fastest-growing actively managed ETF to surpass \$20 billion and that of Mei Ying Sen?
}

\begin{CJK*}{UTF8}{gbsn}
答案：-6.19\%
\end{CJK*}

{\color{blue}
Answer: -6.19\%
}
\\
\end{tcolorbox}
\caption{This example demonstrates the Company Fundamental Reasoning scenario, designed to evaluate a model's basic search capabilities and deep financial mathematical logic capabilities. To resolve the task, the model must demonstrate exceptionally strong abilities in targeted parameter extraction from textual descriptions and precise numerical anchoring within financial charts. It must then construct a compound computational logic to execute complex formulas derived from the integrated multimodal inputs. Finally, the task tests the integrity of the entire reasoning-to-calculation chain, moving from cross-modal data synthesis to higher-order numerical closure.}
\label{fig:secnario-2}
\end{figure*}

\begin{figure*}[htbp] 
\begin{tcolorbox}[
    colback=darkgray!5!white,
    title=Company Fundamental Reasoning, 
    colframe=darkgray!2!darkgray,
    arc=2mm,
    fontupper=\small,
    breakable=false,
    width=\textwidth 
]

\begin{center} 
    \includegraphics[width=0.7\linewidth]{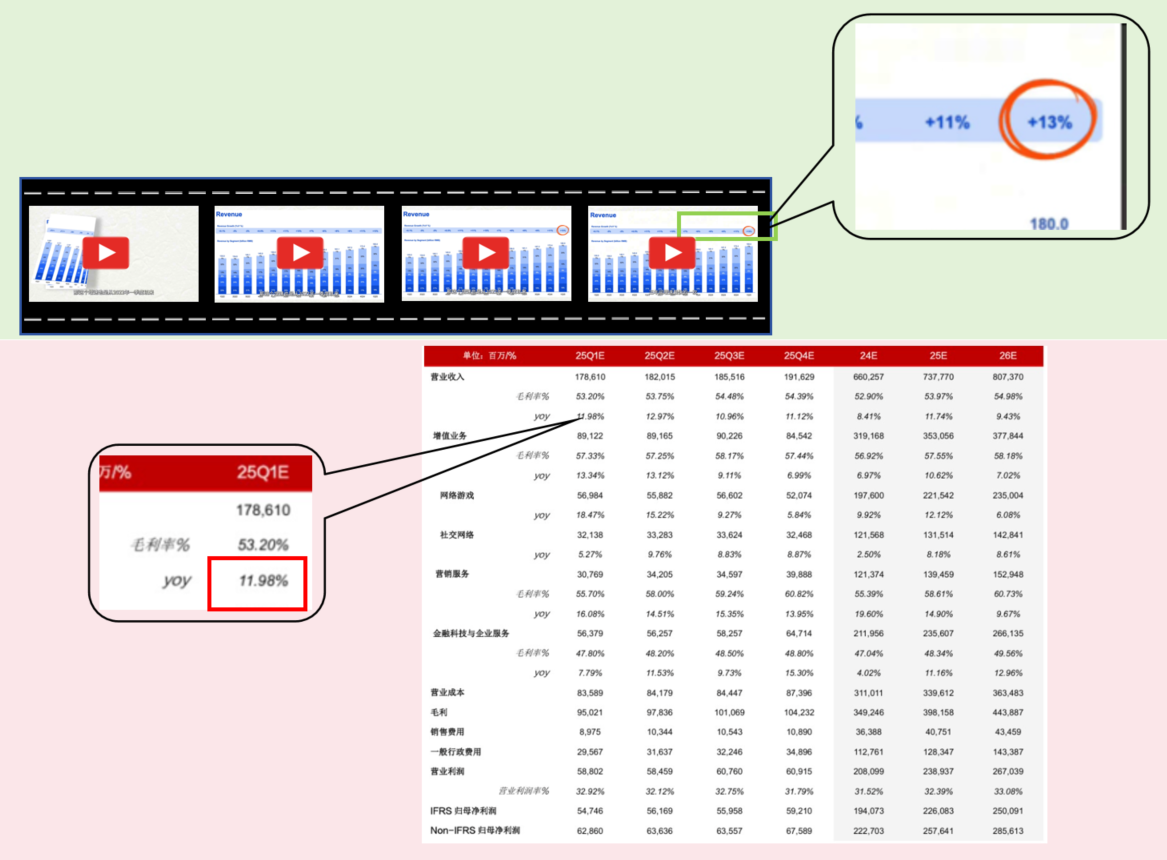}
\end{center}

\begin{CJK*}{UTF8}{gbsn}
问题：腾讯2022年Q1以来收入增速最快的增速最快的一季度增速比2025年Q1营业收入预测同比增速快了多少？（单位：\%，保留两位小数）
\end{CJK*}

{\color{blue}
Question: Since Q1 2022, by how many percentage points did Tencent's fastest quarterly revenue growth rate exceed the predicted year-on-year (YoY) revenue growth rate for Q1 2025? (Unit: \%, rounded to two decimal places)
}

\begin{CJK*}{UTF8}{gbsn}
答案：-1.02\%
\end{CJK*}

{\color{blue}
Answer: -1.02\%
}
\\
\end{tcolorbox}
\caption{This example demonstrates the Company Fundamental Reasoning scenario, designed to evaluate a model's basic search capabilities and deep financial mathematical logic capabilities.}
\label{fig:secnario-2-2}
\end{figure*}

\textbf{Company Fundamental Reasoning:} Building upon basic information extraction, this scenario corresponds to the information alignment and synchronization stage and focuses on interpreting semantic consistencies and information overlaps embedded in textual and graphical content. By rapidly capturing redundancies or correlations across sources, it provides critical alignment-level signals for investment decisions, bridging the gap between heterogeneous data formats and unified business logic.
Data comes from research reports, market commentaries, and analysis materials, where charts no longer directly present explicit financial indicators but convey consistent information through trends, distributions, or relative changes; texts also contain descriptive or summary-based representations.
Figure ~\ref{fig:secnario-2} and Figure ~\ref{fig:secnario-2-2} are the examples of Company Fundamental Reasoning.

\begin{figure*}[htbp] 
\begin{tcolorbox}[
    colback=darkgray!5!white,
    title=Industry Trend Insights, 
    colframe=darkgray!2!darkgray,
    arc=2mm,
    fontupper=\small,
    breakable=false,
    width=\textwidth 
]

\begin{center} 
    \includegraphics[width=0.9\linewidth]{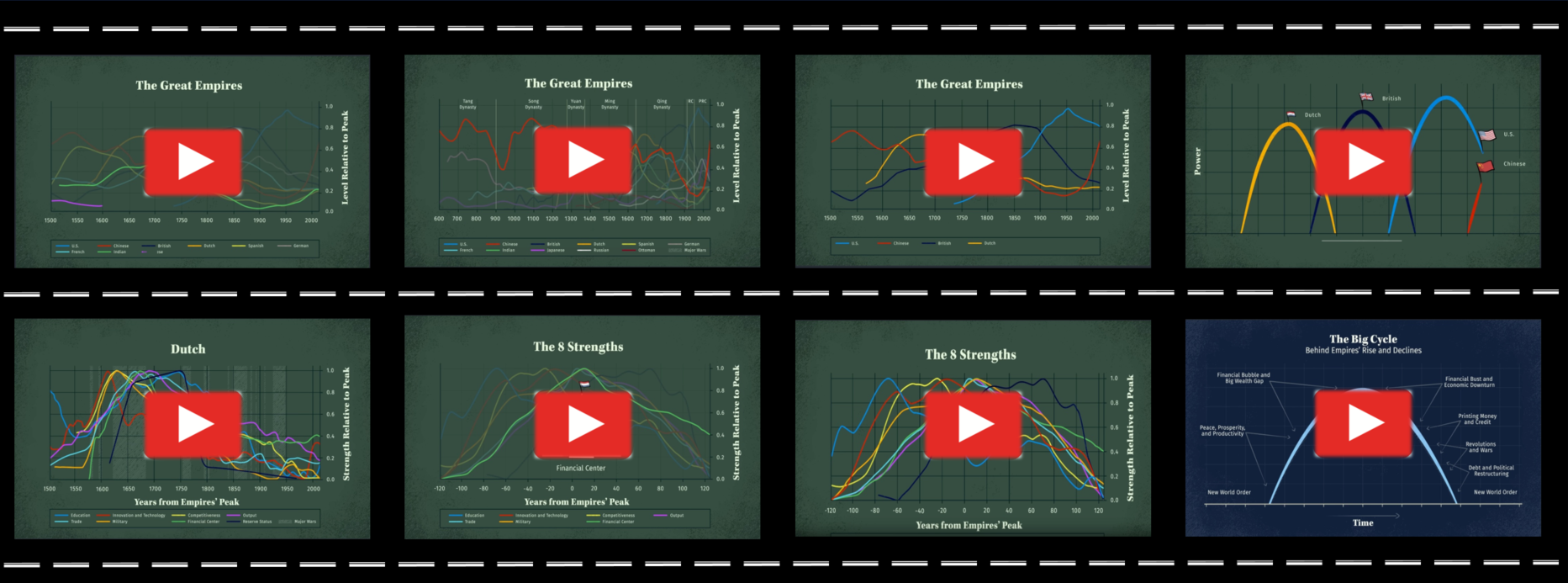}
\end{center}
Text: As I described in my book Principles for Dealing with the Changing World Order, I automated my way looking at the cause effect relationships that are driving both improvements in and worsenings of countries’ conditions so that data is fed into a computer that analyzes it and writes a summary of the current conditions and the long-term prospects for each country.  ... ... The table below shows our aggregate country power gauge and the major drivers, as well as the rank of each measure of power across 24 major countries today and the trajectory over the past twenty years. To understand a country, we start by looking at the big cycles , as well as measures of power that both reflect and drive the rise and fall of a country. While we refer to these factors individually, they are not separate; they interact with and reinforce one another to move a country along its cycle. For the United States, the big cycles look mostly unfavorable. The United States is in an unfavorable position in its economic and financial cycles, with a high debt burden and relatively low expected real growth over the next 10 years (1.3\% per year). The United States has significantly more foreign debts than foreign assets (net IIP is -68\% of GDP). Non-financial debt levels are high (274\% of GDP), and government debt levels are high (127\% of GDP). The bulk (99\%) of these debts are in its own currency, which mitigates its debt risks. The ability to use interest rate cuts to stimulate the economy is low (short rates at 0.1\%), and the country is already printing money to monetize debt. That said, being the world’s leading reserve currency is a large benefit to the US. If this were to change, it would significantly weaken the US position. Internal disorder is a high risk. Wealth, income, and values gaps are large (relative to countries of similar per capita income levels). Regarding Inequality—the top 1\% and top 10\% in the United States capture 19\% and 45\% of income (respectively the 8th and 11th highest share across major countries). ... ..., Half of the measure captures the absolute quantity of educated people at various levels and about half is placed on quality such as higher education rankings, test scores, and average years of education. The US ranks highest in this gauge (driven by strong absolute and relative measures of higher education), with China close behind (due to its large number of educated people). Financial Center:  This gauge measures the level of development and sizes of a country’s financial markets and financial center. We look at absolute measures of transaction shares and market capitalizations, as well as external indices of financial center cities. The US remains the top-ranked power in this metric by a significant margin (driven primarily by its very large share of world equity and debt mar -kets), with China and Europe ranking second and third, respectively. Reserve Currency Status: This gauge measures the extent to which a country’s currency operates as a global reserve currency. We measure reserve currency status by the share of transactions, debts, and central bank reserves that are denominated or held in a country’s currency. Similar to financial center status, the US remains the top-ranked power in this metric by a significant margin, with Europe and Japan ranking second and third, respectively. ... ..., In case it is helpful or interesting to you, you can review those scores below.33 In a few cases where there were no quality measures, I had to create quality measures by adjusting the quantity for a country’s population, turning it into a per capita measure. We did not give reserve currency status scores to the countries that share the euro, which is why those measures are displayed as dashes. 4 Because the notion of competitiveness is inherently relative, we only show the total score for this measure.

{\color{blue}
Question:Synthesizing the analytical logic regarding the 'Big Cycle' and the 'Eight Major Strengths' in the text, as well as the average evolutionary trends of the 'Eight Major Strengths' for all empires before and after their respective peaks as shown in Figure 5, please infer: After an empire's power reaches its peak, which 'strength' indicator declines the fastest (i.e., requires the shortest time to drop from its peak to the 0.2 level)? Please substantiate your argument by combining the trend lines in the video with the implications in the text regarding the vulnerability of 'financial center' status.

Answer: Financial Center
}

\end{tcolorbox}
\caption{This example demonstrates the Industry Trend Insights scenario. It requires the model to extract the evolutionary trends of the 'Eight Major Strengths' indicators for all empires before and after their peaks from the video, focusing on comparing the declining slopes of each indicator curve after the peak, to preliminarily identify that the 'financial center' indicator curve has the steepest declining slope, presenting a trend of rapid decline; secondly, to locate descriptions regarding the 'financial center' indicator in the text, confirming that its core components rely on short-term capital flows and market confidence, and that its strong binding relationship with reserve currency status will trigger a transmission chain of rapid decline when the economic and financial cycle turns unfavorable; and finally, to associate the steep decline trend of the 'financial center' in the video with the high sensitivity and high vulnerability of this indicator in the text, arriving at the conclusion that the 'financial center' indicator declines the fastest after the empire's power reaches its peak.}
\label{fig:secnario-3}
\end{figure*}

\textbf{Industry Trend Insights:} Designed for multi-source data integration scenarios, this task addresses the complexity of deriving core metrics from diverse raw data points found across modalities such as text and charts. By executing precise quantitative formulas across modalities, it prevents computational errors caused by fragmented information and ensures the accuracy and reliability of calculated financial results.
The data includes explicit or implicit links between text and charts, deliberately retaining the incomplete symmetry of expression between modalities, making it impossible for the model to complete the task through surface matching; it must understand the underlying calculation logic behind different modalities. Questions also deepen from single-step parameter extraction to multi-stage formula derivation and sensitivity analysis.
Figure ~\ref{fig:secnario-3} are the examples of Industry Trend Insights.

\begin{figure*}[htbp] 
\begin{tcolorbox}[
    colback=darkgray!5!white,
    title=Financial Risk Sensing, 
    colframe=darkgray!2!darkgray,
    arc=2mm,
    fontupper=\small,
    breakable=false,
    width=\textwidth 
]

\begin{center} 
    \includegraphics[width=0.9\linewidth]{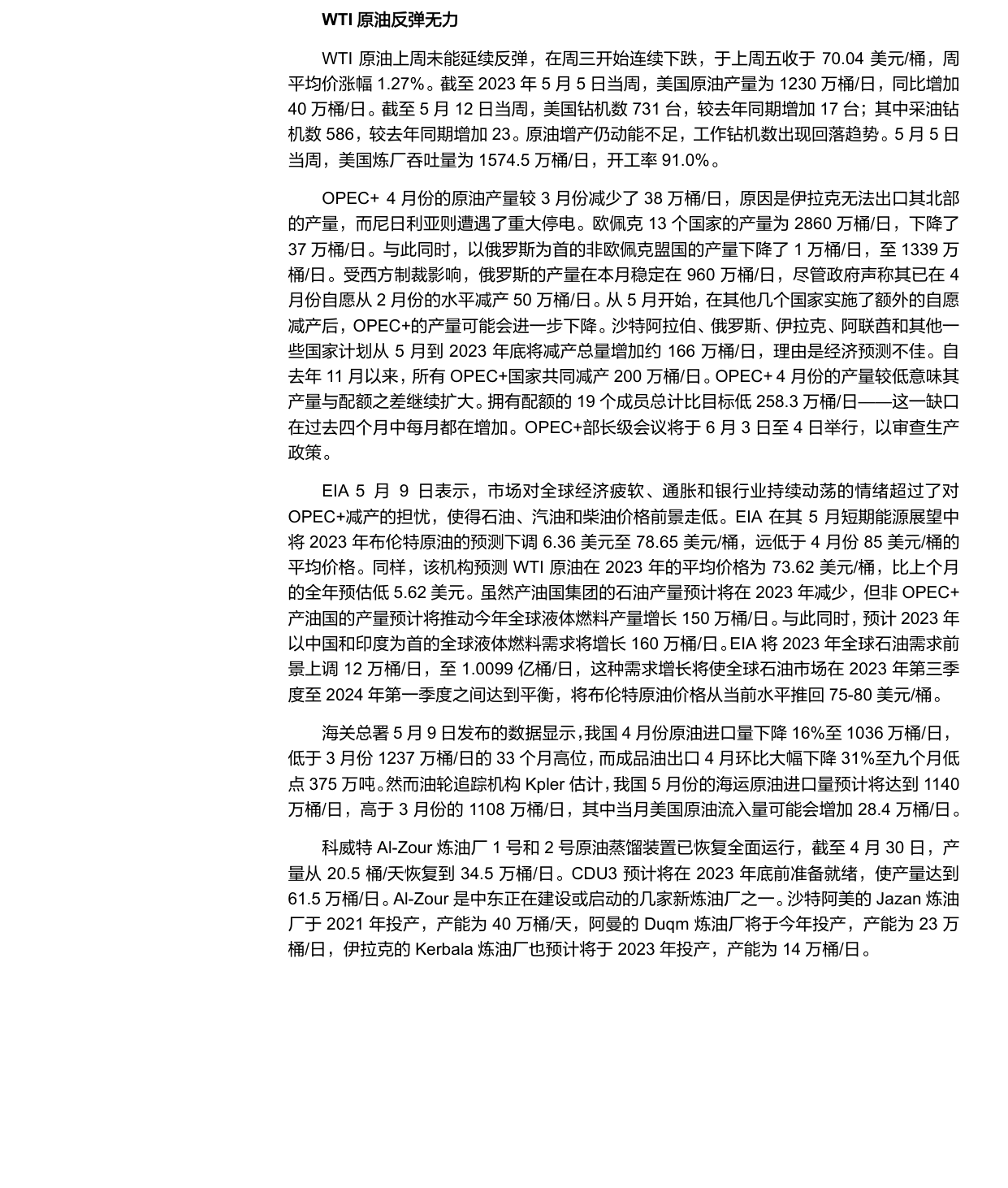}
\end{center}

\begin{CJK*}{UTF8}{gbsn}
问题：结合OPEC+ 4月份实际减产数据、5月到年底的计划减产总量，以及EIA对全球经济情绪的评估，分析EIA下调2023年WTI原油平均价格预测的核心原因，并指出该原因如何导致OPEC+减产对价格的支撑作用被削弱。
\end{CJK*}

{\color{blue}
Question: By combining OPEC+’s actual production cuts implemented in April, the total planned production cuts from May through the end of the year, and the EIA’s assessment of global economic sentiment, analyze the core reason why the EIA lowered its forecast for the average WTI crude oil price in 2023, and explain how this reason weakened the price-supporting effect of OPEC+ production cuts.
}

\begin{CJK*}{UTF8}{gbsn}
答案：市场对全球经济疲软、通胀和银行业动荡的情绪超过了对OPEC+减产的担忧；经济担忧抵消了减产效果，使价格预期下调
\end{CJK*}

{\color{blue}
Answer: Market concerns about global economic weakness, inflation, and turmoil in the banking sector outweighed concerns over OPEC+ production cuts; these economic worries offset the impact of the cuts, leading to a downward revision in price expectations.
}
\\

\end{tcolorbox}
\caption{This example illustrates the Financial Risk Sensing scenario, requiring the model to keenly capture potential downside risk signals. First, it must cross-verify the structural details of OPEC+'s complex production plan with the descriptive macro sentiments highlighted by the EIA, such as global economic weakness, inflation, and banking turmoil. Subsequently, the model must accurately synchronize these heterogeneous signals to identify that market concerns regarding an economic recession have outweighed expectations of support from production cuts. This task evaluates whether the model can logically align these disparate information threads to justify the downward revision of the WTI average price forecast to \$73.62 per barrel.}
\label{fig:secnario-4}
\end{figure*}

\textbf{Financial Risk Sensing:} Corresponding to long-term industry tracking and mid-term analytical assessment, this scenario integrates temporal information from both static text–image inputs and dynamic video data to infer industry development logic, core driving factors, and volatility risks, providing trend-level support for mid-term investment direction selection. 
Question design requires the model to synthesize information changes at different moments to explain and judge industry development trends or market fluctuations. 
Figure ~\ref{fig:secnario-4} are the examples of Financial Risk Sensing.

\begin{figure*}[htbp] 
\begin{tcolorbox}[
    colback=darkgray!5!white,
    title=Asset Allocation Analysis, 
    colframe=darkgray!2!darkgray,
    arc=2mm,
    fontupper=\small,
    breakable=false,
    width=\textwidth 
]

\begin{center} 
    \includegraphics[width=0.9\linewidth]{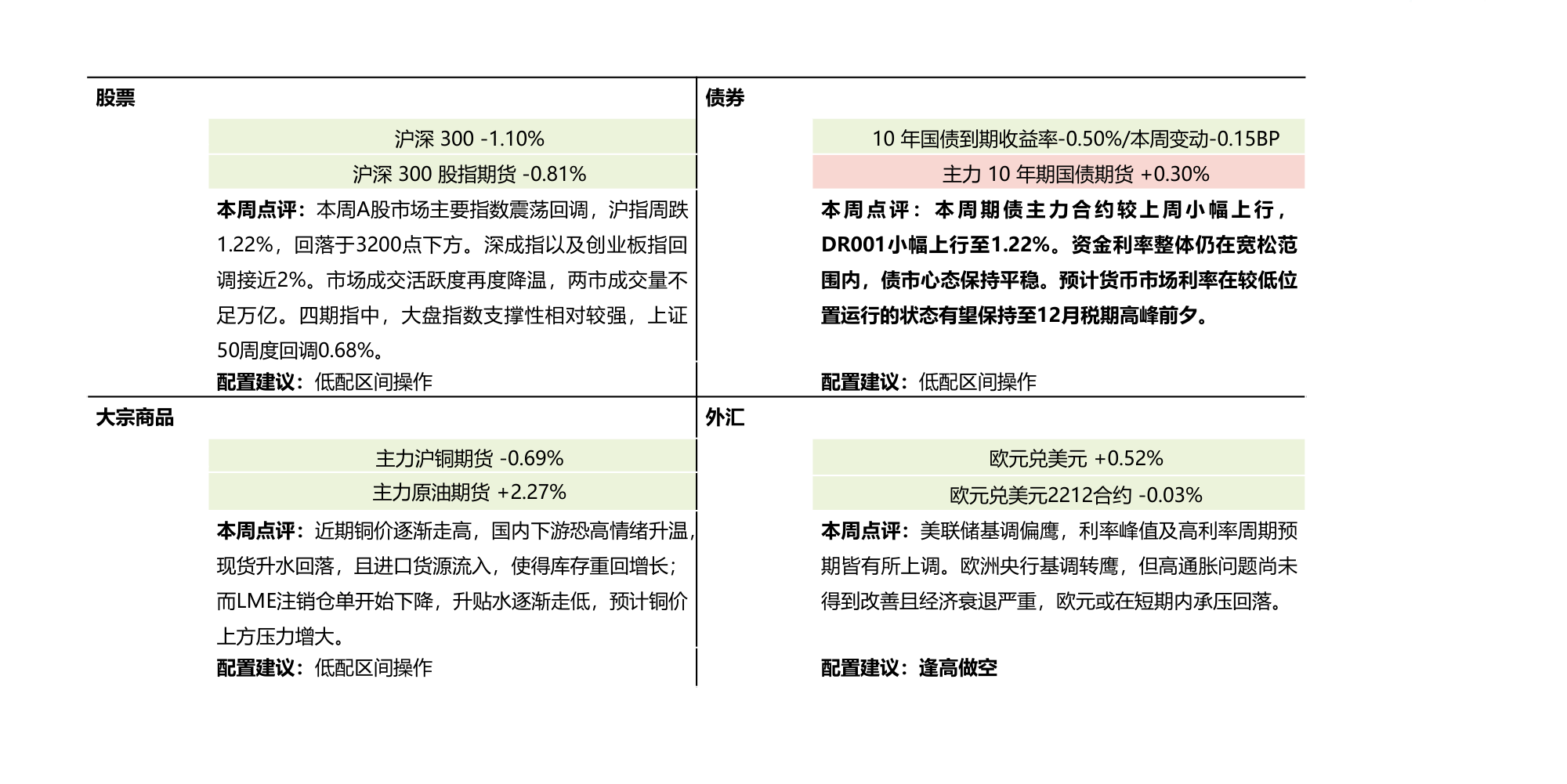}
\end{center}

\begin{CJK*}{UTF8}{gbsn}
问题：基于文字说明中'随后两个交易日内保持日均1000亿元左右的投放规模'的表述，以及图表25所反映的2022年11月市场波动背景，计算从操作当日到后续两个交易日的总投放规模相对于当日净投放额的倍数（保留两位小数）
\end{CJK*}

{\color{blue}
Question: Based on the statement in the text description regarding 'maintaining an average daily injection scale of around 100 billion yuan for the subsequent two trading days,' and the background of market volatility in November 2022 reflected in Figure 25, calculate the multiple of the total injection volume from the day of operation through the subsequent two trading days relative to the net injection amount on that day (rounded to two decimal places).
}

\begin{CJK*}{UTF8}{gbsn}
答案：1.39
\end{CJK*}

{\color{blue}
Answer: 1.39
}
\\
\end{tcolorbox}
\caption{This is a comprehensive question examining macro asset allocation logic and cross-market volatility attribution capabilities. It requires the model to possess keen financial semantic recognition and multi-dimensional indicator alignment capabilities: First, it must extract fluctuation data across categories for the CSI 300 Index, stock index futures, and dominant crude oil futures; second, it needs to accurately calculate the average risk exposure of the equity and energy dual axes in major asset allocation; finally, quantitatively derive the core metric reflecting the true market volatility level, testing the model's quantitative analysis chain from discrete data capturing to composite risk measurement.}
\label{fig:secnario-5}
\end{figure*}

\textbf{Asset Allocation Analysis:} As the ultimate decision-making stage in financial operations, this scenario integrates outputs from all preceding stages and, under multiple constraints such as policy and risk considerations, produces actionable asset allocation strategies. It directly reflects the practical deployment value of the model and supports core investment decision-making. 
Question design requires the model to continuously integrate new information during interaction and maintain decision logic consistency under multimodal, multi-hop reasoning conditions. 
Figure ~\ref{fig:secnario-5} are the examples of Asset Allocation Analysis.

\subsection{Details of Quality Control}
\label{Details of Quality Control}

\textbf{Data Collection}
To ensure authority and real-world alignment, UniFinEval sourced data from global real financial markets, collecting over 2,500 listed company documents and financial research reports, as well as 1,000 financial analysis videos. 
These materials cover both Chinese and English environments across various asset classes, with all sources verified for copyright compliance. 
To maintain high information density, we employed a multi-layered filtering process combining automated scripts with collaborative manual verification to remove low-quality or irrelevant content.

\textbf{Data Filtering}
Experts first used automated scripts to crop pages with financial charts and associated text from PDF reports, forming independent complete images with high information density. For financial analysis videos, scene segmentation, keyframe extraction, and timestamp alignment were performed to ensure semantic alignment between static screenshots and the original video. Subsequently, manual review and screening were conducted, focusing on assessing business representativeness, information effectiveness, and scenario adaptability to fundamentally guarantee data quality and business value.

\textbf{Question Construction}
First, each expert independently completed question design and standard answer annotation based on assigned scenario data to ensure originality and independence, avoiding homogenization of thought. Subsequently, a dual-round cross-validation phase began. Each question was verified by an expert annotator from three dimensions: answer accuracy, logical rigor, and semantic clarity. If explicit errors were found in the question or answer, the question was directly discarded; if doubts existed regarding logic or semantics, it was submitted to a second expert annotator for independent re-verification. If a consensus could not be reached or both annotators were unsure, the data was removed from the benchmark. 

\textbf{Expert Review}
First, all questions had to pass a compliance review, where experts with experience in financial regulatory policy research assessed whether they met financial industry compliance requirements, avoiding sensitive expressions or improper business scenario settings. Second, the expert team conducted a centralized review of business adaptability and difficulty gradients, focusing on checking if questions fit real business flows while ensuring a reasonable difficulty ratio to distinguish model capabilities at different levels in a fine-grained manner. For multi-turn Q\&A tasks, an additional logical coherence review was required, where experts simulated the multi-turn interaction process from the perspective of a real analyst to verify if the dialogue link was smooth, the context consistent, and the decision deduction compliant with business logic.

\subsection{Examples of Environmental Perturbation Simulations}
\label{sec:perturbation}
\begin{figure*}[htbp] 
\begin{tcolorbox}[
    colback=darkgray!5!white,
    title=Stain Simulation, 
    colframe=darkgray!2!darkgray,
    arc=2mm,
    fontupper=\small,
    breakable=false,
    width=\textwidth 
]

\begin{center} 
    \includegraphics[width=0.9\linewidth]{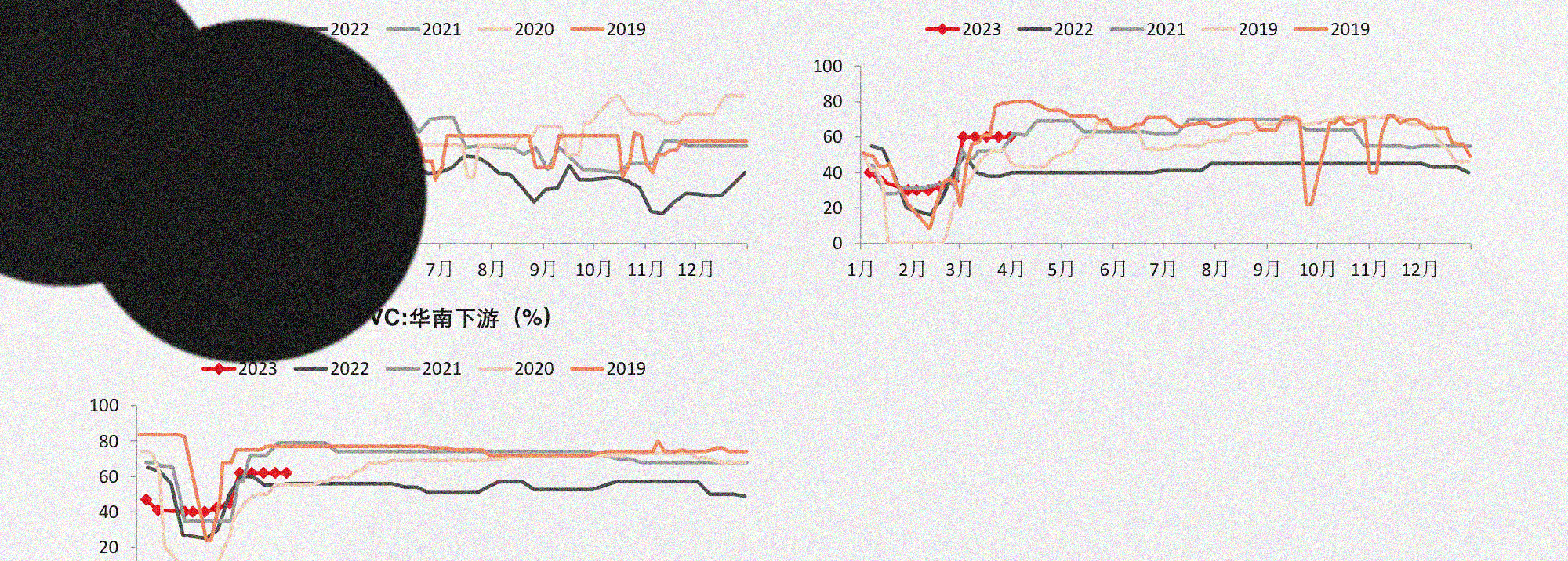}
\end{center}

\begin{CJK*}{UTF8}{gbsn}
问题：基于提供的金融研报图像（包含左上角、右上角和底部三个子图，均展示PVC华南下游开工率（\%），请分析：在右上角子图中，2019年开工率在2月的下降趋势与2023年在3月的上升趋势相比，哪一段的波动更剧烈？
\end{CJK*}

{\color{blue}
Question: Based on the provided financial research report image (which contains three subplots located at the top-left, top-right, and bottom, all showing the PVC South China downstream operating rate (\%), please analyze the following: In the top-right subplot, when comparing the downward trend in the operating rate in February 2019 with the upward trend in March 2023, which period exhibits more pronounced volatility?
}

\begin{CJK*}{UTF8}{gbsn}
答案：2019年2月的下降段更剧烈
\end{CJK*}

{\color{blue}
Answer: The decline in February 2019 was more pronounced.
}
\\
\end{tcolorbox}
\caption{This is an example of stain simulation perturbation, simulating ink smudges, coffee stains, or mold spots acquired by financial documents during circulation or storage.}
\label{fig:per-1}
\end{figure*}

In real-world financial office and auditing scenarios, paper-based documents often suffer from degraded visual quality due to improper storage, suboptimal photographing angles, or physical wear and tear. To more comprehensively evaluate model robustness under extreme real-world conditions, UniFinEval specifically introduces three environment-noise processing tasks targeting physical entities. These simulations are designed to mimic common degradation phenomena encountered during the digitization of financial receipts and reports. The categories and definitions are as follows: 

\textbf{Stain Simulation:} This task simulates ink smudges, coffee stains, or mold spots that may occur during the circulation or storage of financial documents. Such disturbances are implemented by randomly overlaying masks of varying shapes and transparency on the images, partially occluding text or accounting relationships. This setting tests the model’s ability to recover incomplete information and its robustness to visual interference. Figure ~\ref{fig:per-1} provides an example of this case.

\begin{figure*}[htbp] 
\begin{tcolorbox}[
    colback=darkgray!5!white,
    title=Perspective Simulation, 
    colframe=darkgray!2!darkgray,
    arc=2mm,
    fontupper=\small,
    breakable=false,
    width=\textwidth 
]

\begin{center} 
    \includegraphics[width=0.9\linewidth]{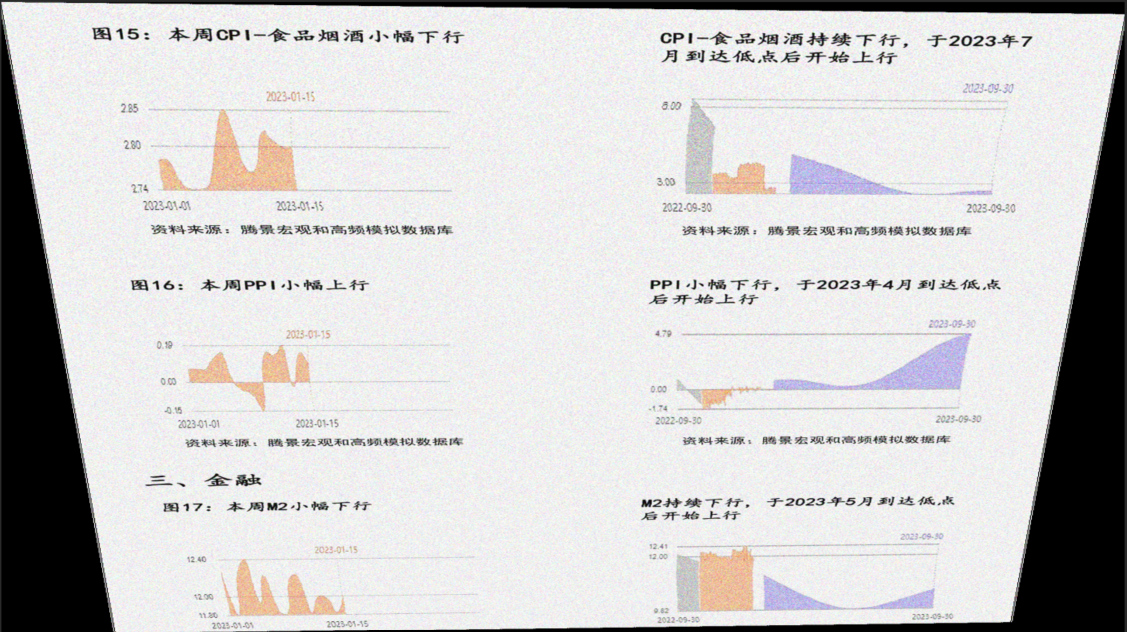}
\end{center}

\begin{CJK*}{UTF8}{gbsn}
问题：综合图15、16、17右侧子图对PPI、M2、CPI-食品烟酒三个指标触底反弹时间的描述（PPI于2023年4月、M2于2023年5月、CPI-食品烟酒于2023年7月到达低点后开始上行），以及左侧子图显示的2023年1月15日当周三个指标均小幅下行的特征，判断在2023年1月15日时点，哪个指标的下行期剩余时间最短？需严格依据图像文字信息进行跨指标时序比较和趋势阶段定位。
\end{CJK*}

{\color{blue}
Question: Based on a comprehensive analysis of the descriptions in the right-hand subplots of Figures 15, 16, and 17 regarding the bottoming-out and rebound timing of three indicators—PPI, M2, and CPI-Food, Tobacco, and Liquor (specifically, PPI reached its low point and began to rise in April 2023, M2 in May 2023, and CPI-Food, Tobacco, and Liquor in July 2023)—combined with the characteristic shown in the left-hand subplots where all three indicators experienced a slight decline during the week of January 15, 2023, determine which indicator had the shortest remaining time in its downward phase as of the time point of January 15, 2023. You are required to strictly base your analysis on the text information within the images to perform a cross-indicator temporal comparison and trend phase positioning.
}

\begin{CJK*}{UTF8}{gbsn}
答案：PPI
\end{CJK*}

{\color{blue}
Answer: PPI
}
\\
\end{tcolorbox}
\caption{This is an example of perspective simulation perturbation, simulating the geometric distortion caused by capturing images with mobile phones or scanners at non-parallel angles.}
\label{fig:per-2}
\end{figure*}

\textbf{Perspective Simulation:} This task simulates geometric distortions caused by photographing documents with mobile phones or scanners from non-parallel angles. By applying perspective projection transformations to the original images, documents exhibit trapezoidal warping or edge stretching. This requires the model to possess strong spatial perception capabilities in order to accurately extract structured data under coordinate shifts. Figure ~\ref{fig:per-2} demonstrates this perturbation type.

\textbf{Crease Simulation:} This task simulates physical creases left on paper documents after being folded or crumpled. Creases introduce localized linear highlights or shadows and may cause text strokes to break or become misaligned. Such simulations effectively assess the OCR accuracy of multimodal models when dealing with non-flat document layouts. 

\textbf{Curvature Simulation:} Simulates the non-linear curling effects presented by thick financial reports, book binding areas, or unflattened paper. Unlike perspective distortion, this type of interference causes text lines to distort irregularly following the curvature of the paper, focusing on testing the model's capabilities in layout analysis, dewarping, and character rectification under complex 3D deformations. 

\textbf{Noise Simulation:} Simulates image background noise resulting from aging scanner sensors, transmission compression loss, or low-light shooting environments. By injecting Gaussian noise or salt-and-pepper noise into the original image to reduce the Signal-to-Noise Ratio (SNR), it aims to evaluate the model's ability to accurately recognize fine details of numbers and text under low-quality imaging conditions characterized by blurriness and strong graininess. 

\section{Details of MLLMs}
\label{modeloverview}
We list details of the MLLMs evaluated using UniFinEval in Table~\ref{tab:model-overview}.

\begin{table*}[!htbp]
\centering
\caption{Models evaluated in this paper. The "Access" column shows whether we have full access to the model weights or we can only access through API. The “Version Date” column shows the release date of the corresponding version of the model we evaluated.}
\label{tab:model-overview}
\vspace{10pt}
\resizebox{0.9 \textwidth}{!}{
\begin{tabular}{llcccc}
    \toprule[2pt]
    \textbf{Category} & \textbf{Model} & \textbf{Creator} & \textbf{Parameter} & \textbf{Access} & \textbf{Version Date}\\
    \midrule[1pt]
    Close-Source 
    & Gemini-3-pro-preview & Google & Undisclosed & API & 2025.11\\
    & GPT-5.1-2025-11-13 & OpenAI & Undisclosed & API & 2025.11\\
    & Grok-4.1-Fast-reasoning  & xAI & Undisclosed & API & 2025.11 \\
    & Claude-Sonnet-4.5-20250929 & Anthropic & Undisclosed & API & 2025.9\\
    \midrule[1pt]    
    Open-Source 
    & Qwen3-VL-235B-A22B-thinking & Alibaba Cloud & 235B & Weights & 2025.11 \\
    & Qwen3-VL-32B-thinking & Alibaba Cloud & 32B & Weights & 2025.11\\
    & InternVL3.5-241B-A28B & Shanghai AI Lab & 241B & Weights & 2025.8\\
    & InternVL3.5-30B-A3B & Shanghai AI Lab & 30B & Weights & 2025.8 \\
    & Llama-3.2-11B-Vision & Meta AI & 11B & Weights & 2024.9\\
    & MiniCPM-V-4.5 & OpenBMB & 9B & Weights & 2025.9 \\

    \bottomrule[2pt]
\end{tabular}}
\end{table*}

\subsection{Examples for Error Analysis}
\label{error analysis}
Building upon the error categories introduced in the main text, this appendix provides a qualitative explanation of each error type. The purpose is to clarify the characteristics and underlying causes of common failure modes exhibited by MLLMs in financial reasoning tasks, rather than to introduce additional experimental analyses.

\textbf{Financial Image Perception and Data Interpretation Errors:}This category of errors refers to failures occurring at the visual perception and data interpretation stage of financial multimodal reasoning. In finance-related tasks, models are often required to accurately extract numerical values, trends, and structural relationships from visual inputs such as financial charts, tables, trend plots, and annotated figures. Errors arise when the model misidentifies key visual elements (e.g., axes, legends, data points) or incorrectly interprets visually encoded information. These issues are particularly prominent under conditions of high visual complexity, noise, dense annotations, low image resolution, or overlapping graphical components, and they frequently propagate to subsequent reasoning steps, resulting in incorrect financial conclusions.

\textbf{Financial Knowledge Reasoning and Domain-Specific Understanding Errors:}This type of error stems from limitations in the model’s understanding of financial concepts, professional terminology, and domain-specific reasoning principles. Financial tasks are characterized by high knowledge density and strong reliance on specialized concepts, such as financial ratios, capital structure, accounting rules, and risk assessment frameworks. Models may misinterpret these concepts, confuse related financial terms, or apply inappropriate domain logic, even when the input information is correctly perceived. As a result, the generated reasoning may deviate from established financial analysis practices, leading to conclusions that are logically coherent in form but flawed from a professional finance perspective.

\textbf{Financial Computation and Numerical Analysis Errors:}Financial computation and numerical analysis errors occur when models fail to produce accurate quantitative results during multi-step numerical reasoning. In finance, precise calculations are essential for tasks involving valuation, profitability analysis, investment returns, or comparative financial metrics. Models may make arithmetic mistakes, apply incorrect calculation sequences, suffer from rounding or precision loss, or fail to consistently track intermediate numerical states. Such errors indicate limitations in the reliability of numerical reasoning, where even minor computational inaccuracies can lead to substantially distorted financial interpretations.

\textbf{Cross-modal Data Integration and Alignment Errors:}Cross-modal data integration and alignment errors arise when models fail to correctly associate and fuse information across different input modalities, such as text, images, and tables. Financial multimodal tasks often require joint reasoning over heterogeneous sources, for example aligning textual descriptions with chart-based evidence or tabular financial disclosures. Errors occur when the model overlooks critical cues, mismatches references between modalities, or incorrectly prioritizes information from one modality while neglecting others. These misalignments can result in incomplete or inconsistent interpretations of the overall financial context, ultimately undermining the accuracy of the reasoning process.

\textbf{Inconsistent Financial Reasoning and Hallucination Errors:}This category captures errors related to logical inconsistency and hallucinated content in financial reasoning. Such errors occur when models generate conclusions that are not supported by the provided evidence, contradict earlier reasoning steps, or rely on fabricated assumptions. In financial scenarios, this may manifest as unfounded predictions of market trends, speculative assessments of corporate financial health, or confident but unsupported explanations of financial outcomes. These errors are particularly problematic in decision-sensitive financial applications, as they may present misleadingly plausible narratives that lack factual or logical grounding.

\begin{figure*}[htbp] 
\begin{tcolorbox}[
    colback=darkgray!5!white,
    title=Financial Image Perception and Data Interpretation Errors, 
    colframe=darkgray!2!darkgray,
    arc=2mm,
    fontupper=\small,
    breakable=false,
    width=\textwidth 
]

\begin{center} 
    \includegraphics[width=0.8\linewidth]{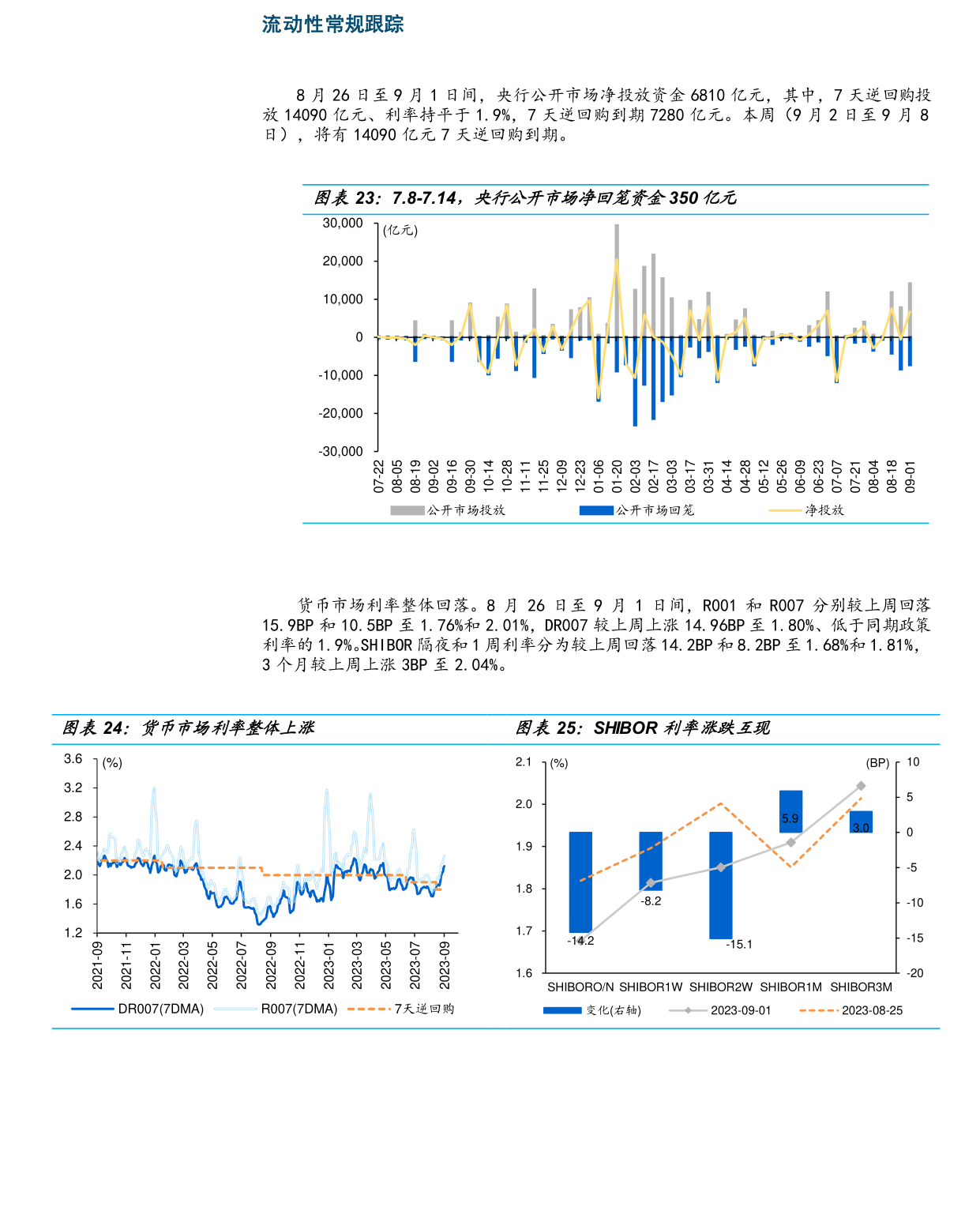}
\end{center}

\begin{CJK*}{UTF8}{gbsn}
问题：结合文字说明中SHIBOR 3个月利率的周度变化数据与图表25中SHIBOR 1个月利率的周度变化数据，计算3个月与1个月SHIBOR利差在8月26日至9月1日期间的基点变化值（需考虑利差定义方向）。
\end{CJK*}

{\color{blue}
Question: Based on the weekly change data of the 3-month SHIBOR rate provided in the textual description and the weekly change data of the 1-month SHIBOR rate shown in Figure 25, calculate the basis-point change in the SHIBOR spread between the 3-month and 1-month tenors during the period from August 26 to September 1 (taking into account the definition and direction of the spread).
}

\begin{CJK*}{UTF8}{gbsn}
答案：-2.9

{\color{blue}
Answer: -2.9
}

模型输出:3个月与1个月SHIBOR利差在8月26日至9月1日期间的基点变化值为7.0 BP。

{\color{blue}
Model Output:The change in the basis-point spread between the 3-month and 1-month SHIBOR over the period from August 26 to September 1 is 7.0 bps.
}

\end{CJK*}
\end{tcolorbox}
\caption{This case represents a typical example of Financial Image Perception and Data Interpretation Errors, where the model incorrectly reads the SHIBOR data in Figure 25 by misidentifying the SHIBOR 1-month value on August 26 as 19.2 bps (instead of the correct value), which subsequently leads to an erroneous spread calculation. This error highlights the model’s limitations in accurately perceiving and interpreting key numerical information from financial charts, particularly when dealing with visually complex financial figures, where visual misperception can easily propagate into downstream reasoning errors.}
\label{fig:error1}
\end{figure*}

\begin{figure*}[htbp] 
\begin{tcolorbox}[
    colback=darkgray!5!white,
    title=Financial Knowledge Reasoning and Domain-Specific Understanding Errors, 
    colframe=darkgray!2!darkgray,
    arc=2mm,
    fontupper=\small,
    breakable=false,
    width=\textwidth 
]

\begin{center} 
    \includegraphics[width=0.9\linewidth]{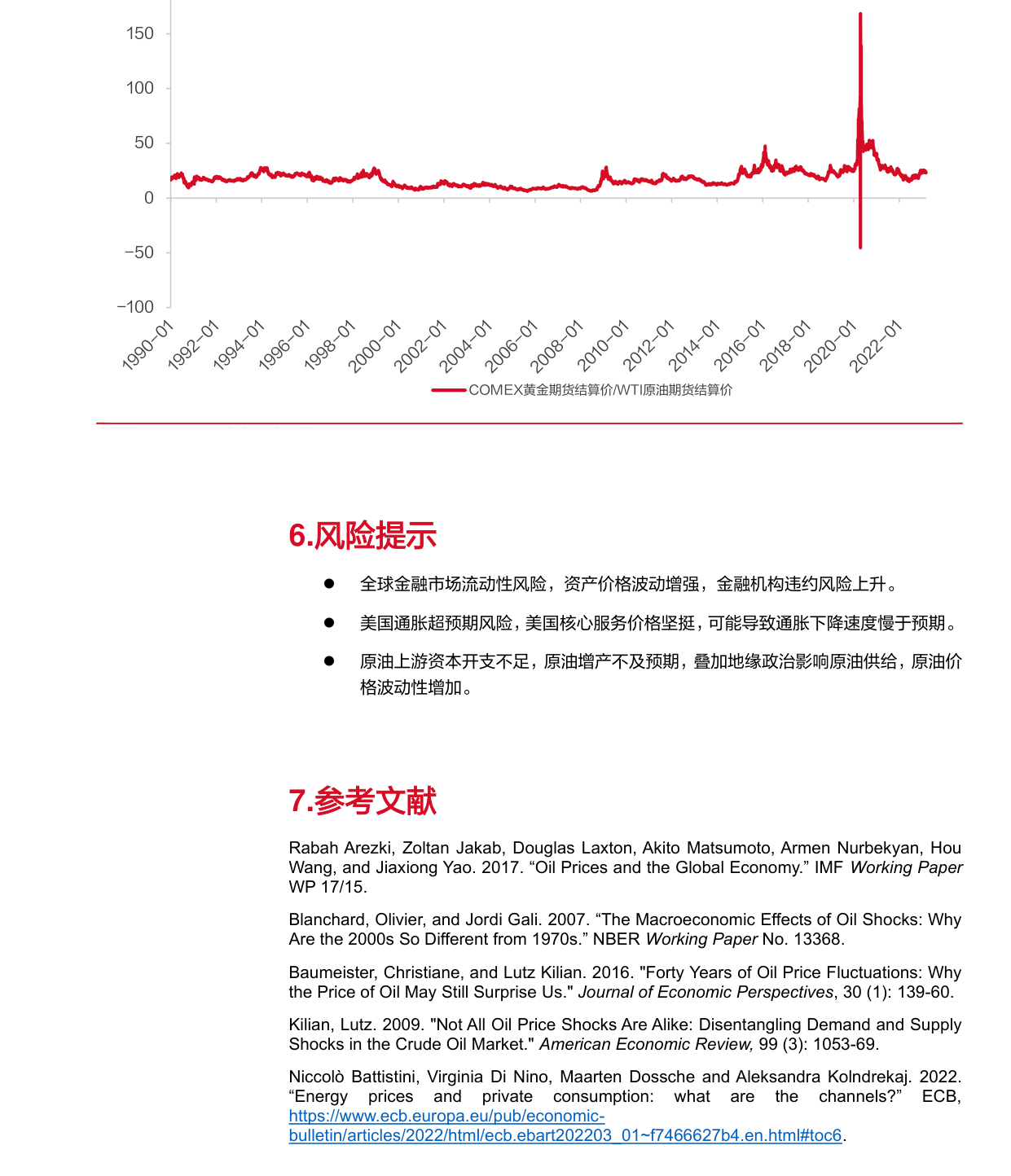}
\end{center}

\begin{CJK*}{UTF8}{gbsn}
问题：折线图中2020年COMEX黄金期货结算价/WTI原油期货结算价出现极端峰值，结合风险提示内容，该现象最直接印证了哪一风险传导路径的结果？请从风险提示中提取核心结果短语作为答案。
\end{CJK*}

{\color{blue}
Question: The line chart shows an extreme spike in the 2020 ratio of COMEX gold futures settlement prices to WTI crude oil futures settlement prices. In conjunction with the risk warning section, which risk transmission pathway does this phenomenon most directly confirm? Please extract the core conclusion phrase from the risk warning as your answer.
}

\begin{CJK*}{UTF8}{gbsn}
答案：金融市场流动性结果
\end{CJK*}

{\color{blue}
Answer: Financial market liquidity outcome
}

\begin{CJK*}{UTF8}{gbsn}

模型输出：原油价格波动性增加
\end{CJK*}

{\color{blue}
Model Output:An increase in crude oil price volatility.
}

\end{tcolorbox}
\caption{This case exemplifies a Financial Knowledge Reasoning and Domain-Specific Understanding Error, where the model incorrectly attributes the extreme peak observed in the chart to an increase in crude oil price volatility, whereas the ground-truth explanation emphasizes financial market liquidity conditions. This error reflects the model’s insufficient understanding of domain-specific financial concepts, as it fails to correctly identify the core drivers underlying market fluctuations—such as liquidity risk—thereby leading to reasoning that deviates from the key financial semantics.}
\label{fig:error2}
\end{figure*}

\begin{figure*}[htbp] 
\begin{tcolorbox}[
    colback=darkgray!5!white,
    title=Financial Computation and Numerical Analysis Errors, 
    colframe=darkgray!2!darkgray,
    arc=2mm,
    fontupper=\small,
    breakable=false,
    width=\textwidth 
]

\begin{center} 
    \includegraphics[width=0.7\linewidth]{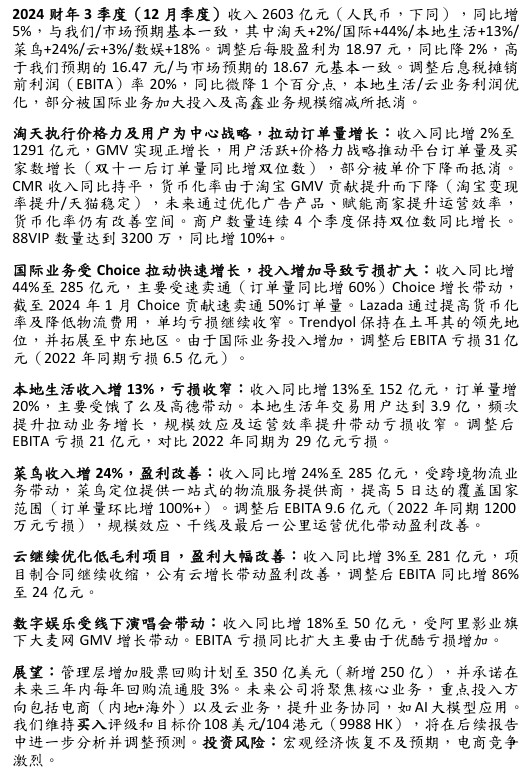}
\end{center}

\begin{CJK*}{UTF8}{gbsn}
问题：阿里巴巴2024年4季度收入占比增长最高的子集团相对2023财年同比增长率为多少？（保留两位小数，单位\%）
\end{CJK*}

{\color{blue}
Question: What is the year-on-year growth rate (relative to fiscal year 2023) of the subsidiary group with the highest increase in revenue share in Alibaba’s fourth quarter of 2024? (Round to two decimal places; unit: \%).
}

\begin{CJK*}{UTF8}{gbsn}
答案：7.60\%
\end{CJK*}

{\color{blue}
Answer: 7.60\%
}

\begin{CJK*}{UTF8}{gbsn}

模型输出：52.05\%
\end{CJK*}

{\color{blue}
Model Output: 52.05\%}

\end{tcolorbox}
\caption{This example illustrates an error where the model computes overall revenue share growth instead of the year-over-year growth rate of the subgroup with the largest increase in revenue share, as required by the question. Although the year-over-year growth rates of individual subgroups are correctly calculated, the model fails to derive revenue share growth based on changes in revenue proportions, resulting in a discrepancy between the predicted value (52.05\%) and the ground-truth answer (7.60\%).
Note: The original question contains nine images; only the first image is shown.}
\label{fig:error3}
\end{figure*}

\begin{figure*}[htbp] 
\begin{tcolorbox}[
    colback=darkgray!5!white,
    title=Cross-modal Data Integration and Alignment Errors in Finance, 
    colframe=darkgray!2!darkgray,
    arc=2mm,
    fontupper=\small,
    breakable=false,
    width=\textwidth 
]

\begin{center} 
    \includegraphics[width=0.7\linewidth]{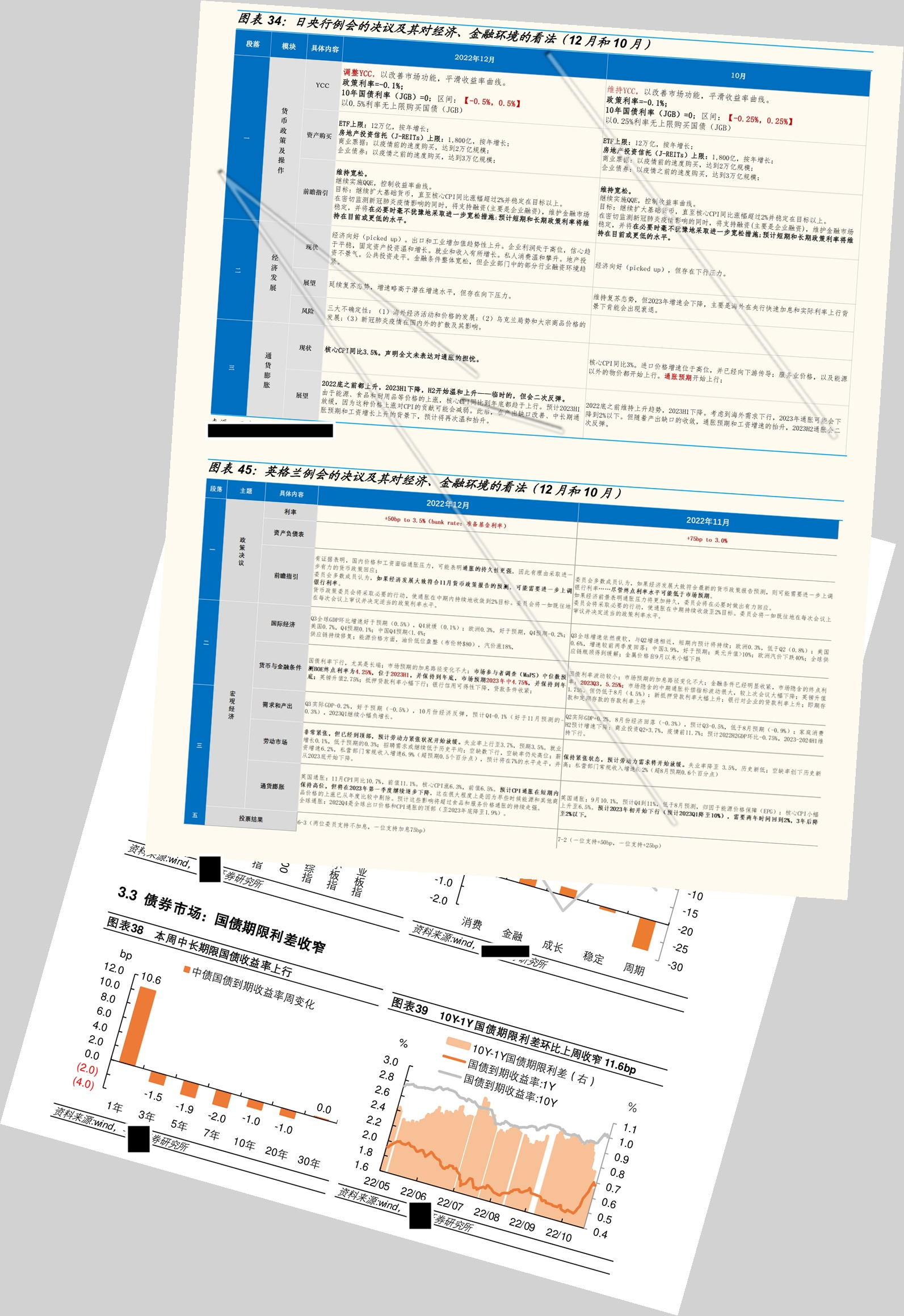}
\end{center}

Question: As the chief strategist of a global macro hedge fund, when constructing a transmission model of overseas central bank policies to domestic short-term interest rates, it is necessary to strictly rely on explicit textual data and conduct multi-step verification. Please compare the statement regarding “a possible secondary rebound” in the “Inflation – Outlook” row of the Bank of Japan’s December 2022 policy meeting table (Figure 34) with the weekly change in the 1-year Chinese government bond yield shown in the ChinaBond market chart (Figure 38). It is known that the Bank of England’s October policy meeting (Figure 45) explicitly stated that inflation “will take two years to return to 2\%,” and Figure 39 shows that the 10Y–1Y government bond term spread narrowed by 11.6 bp month-on-month. Given the model assumption that short-term inflation expectations dominate movements in the 1-year yield, and that “secondary rebound” specifically refers to the short-term risk of an inflation pickup in the second half of 2023, please deduce: which specific numerical change in the ChinaBond market is logically and directly supported by the Bank of Japan’s statement? Please answer by citing the explicit textual content specified in the question prompt.

Answer: 10.6 bp

Model Output: The positive change in the weekly variation of the 1-year government bond yield in the ChinaBond market chart (Figure 38).

\end{tcolorbox}
\caption{This example illustrates a failure in cross-modal alignment, where the model does not effectively align the numerical information in the chart with the logical requirements of the text. Instead of extracting the specific value (10.6 bps) from Figure 38, the model merely describes the qualitative property of the change (a positive variation). This reflects limitations in cross-modal data integration and alignment, as the model fails to precisely associate textual reasoning with chart-based numerical evidence, resulting in the loss of critical financial information.}
\label{fig:error4}
\end{figure*}

\begin{figure*}[htbp] 
\begin{tcolorbox}[
    colback=darkgray!5!white,
    title=Inconsistent Financial Reasoning and Hallucination Errors, 
    colframe=darkgray!2!darkgray,
    arc=2mm,
    fontupper=\small,
    breakable=false,
    width=\textwidth 
]

\begin{center} 
    \includegraphics[width=0.9\linewidth]{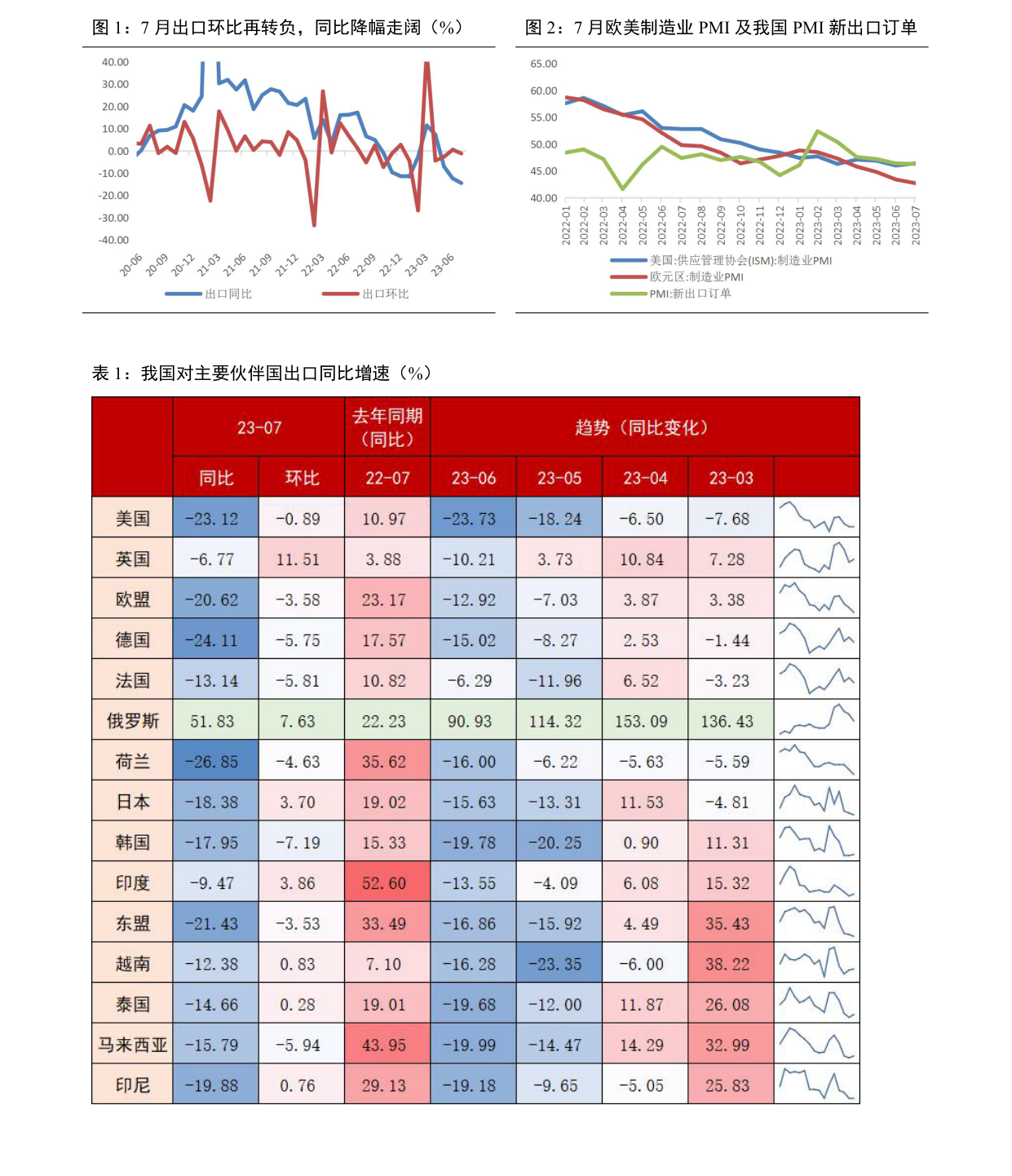}
\end{center}

\begin{CJK*}{UTF8}{gbsn}
问题：基于表1计算23-07与23-06出口同比增速的变化值（23-07同比减23-06同比），并筛选出该变化值小于-5\%的欧美主要伙伴国（限定为美国、欧盟、英国、德国、法国）。结合图2中2023年上半年（23-01至23-07）欧美制造业PMI的持续下行趋势，判断出口同比恶化幅度最大的国家/地区是哪个？
\end{CJK*}

{\color{blue}
Question: Based on Table 1, calculate the change in the year-on-year export growth rate between July 2023 and June 2023 (July 2023 YoY minus June 2023 YoY), and identify the major Western partner economies for which this change is less than -5\% (limited to the United States, the European Union, the United Kingdom, Germany, and France). In conjunction with the sustained downward trend in manufacturing PMI in Europe and the United States during the first half of 2023 (January–July 2023) shown in Figure 2, determine which country/region experienced the largest deterioration in year-on-year export growth.
}

\begin{CJK*}{UTF8}{gbsn}
答案：德国
\end{CJK*}

{\color{blue}
Answer: Germany
}

\begin{CJK*}{UTF8}{gbsn}
模型输出：欧盟, 德国, 法国
\end{CJK*}

{\color{blue}
Model Output: The European Union, Germany, and France
}

\end{tcolorbox}
\caption{This example demonstrates an inconsistency and hallucination issue, where the model generates multiple answers (the European Union, Germany, and France) despite the question requiring a single correct response. Although the correct answer is included, the model introduces irrelevant information, violating the single-answer constraint. This behavior reflects deficiencies in consistent financial reasoning and hallucination control, as the model fails to precisely apply the filtering criteria to produce a unique and accurate output, resulting in redundant and imprecise answers.}
\label{fig:error5}
\end{figure*}

Representative erroneous examples for all five error categories are illustrated in Figures~\ref{fig:error1}--\ref{fig:error5}.

\section{Prompts Used in This Study}
\label{prompt}
We provide representative prompt examples for evaluation. Specifically, the prompt examples for evaluation are shown in Table ~\ref{tab:evaluation1}, Table ~\ref{tab:evaluation2}.


\begin{figure*}[htbp] 
\begin{tcolorbox}[
    colback=darkgray!5!white,
    title=Financial Risk Sensing, 
    colframe=darkgray!2!darkgray,
    arc=2mm,
    fontupper=\small,
    breakable=false,
    width=\textwidth 
]

\begin{center} 
    \includegraphics[width=0.9\linewidth]{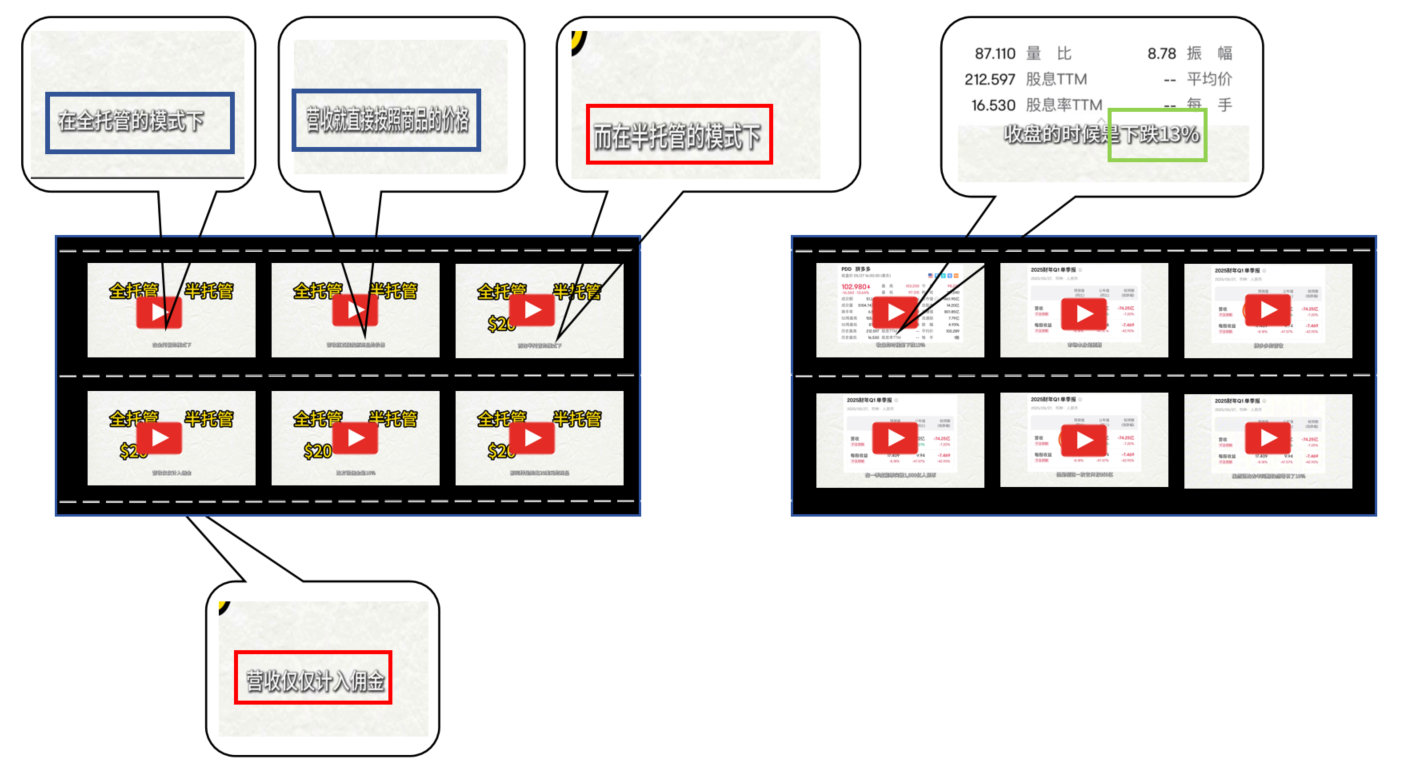}
\end{center}

\begin{CJK*}{UTF8}{gbsn}
问题：请问拼多多2025年Q1营业收入波动的主要原因是什么？
\end{CJK*}

{\color{blue}
Question: What are the primary reasons for Pinduoduo's revenue fluctuations in Q1 2025?
}

\begin{CJK*}{UTF8}{gbsn}
答案：拼多多商业模式由全托管转为半托管。
\end{CJK*}

{\color{blue}
Answer: Pinduoduo's (Temu) Business Model Transition from Fully Managed to Semi-Managed.
}
\\

\end{tcolorbox}
\caption{This example illustrates the Financial Risk Sensing scenario. This task requires the model to reason across two videos to identify the primary factors driving revenue fluctuations and to further recognize the underlying potential risks.
}
\label{fig:secnario-4-2}
\end{figure*}


\begin{figure*}[htbp] 
\begin{tcolorbox}[
    colback=darkgray!5!white,
    title=Cross-modal Data Integration and Alignment Errors in Finance:Text–Image Alignment Example, 
    colframe=darkgray!2!darkgray,
    arc=2mm,
    fontupper=\small,
    breakable=false,
    width=0.9\textwidth 
]

\begin{center} 
    \includegraphics[width=0.8\linewidth]{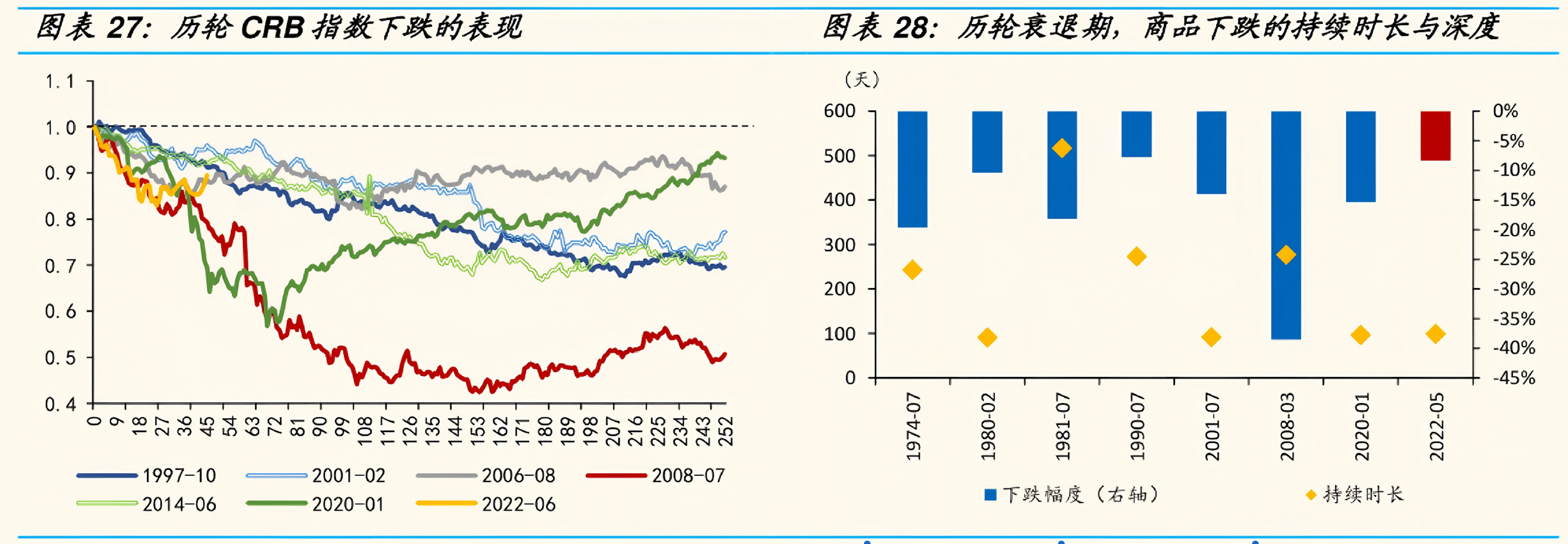}
\end{center}

\begin{CJK*}{UTF8}{gbsn}
文本：10 月以来，资本市场呈现系列新变化，美国股债 “跷跷板” 效应再现，其他子市场亦显露转变，意味着美联储加息周期下的紧缩交易影响已近消退 —— 正常经济周期中股债跷跷板本为常态，仅流动性环境显著变化时易出现同涨同跌，而年初多数时间段的该类情形已逐步缓解，且市场对加息终止的反应较紧缩预期更为敏感，...,资产配置方面，美债与贵金属已成为优选，不过短期交易行为或仍对债市形成干扰，而全球需求走弱背景下，商品市场因与产出缺口显著相关，表现或将受到压制；反观历史，衰退前后黄金、美债等避险资产表现往往较好，当下其配置价值或随尾部风险暴露进一步凸显。
\end{CJK*}

{\color{blue}
Text:Since October, capital markets have exhibited a series of new developments. The U.S. equity–bond “seesaw” effect has re-emerged, and shifts have also become evident across other sub-markets, indicating that the impact of tightening trades under the Federal Reserve’s rate-hiking cycle is nearing dissipation — in a normal economic cycle, the equity–bond seesaw is the norm, with simultaneous rises or declines typically occurring only when liquidity conditions change significantly. The frequent co-movement observed during much of the early part of the year has gradually eased, and markets appear to be more sensitive to signals of an end to rate hikes than to tightening expectations; …, from an asset-allocation perspective, U.S. Treasuries and precious metals have emerged as preferred choices. Nevertheless, short-term trading activity may still interfere with the bond market, while against the backdrop of weakening global demand, commodity markets—given their strong correlation with the output gap—are likely to remain under pressure. Looking back at history, safe-haven assets such as gold and U.S. Treasuries have often performed well around recessionary periods, and under current conditions, their allocation value may become increasingly pronounced as tail risks continue to surface.
}

\begin{CJK*}{UTF8}{gbsn}
问题:结合研究发现第4点中'当下'的时效性指代、图表28的衰退期年份序列，以及风险提示第2点对经济衰退类型的描述，推断与当前商品市场表现直接关联的衰退期具体年份。
\end{CJK*}

{\color{blue}
Question: 
Based on the temporal reference of ‘the current period’ in Research Finding 4, the sequence of recession years shown in Figure 28, and the description of the recession type in Risk Warning 2, infer the specific recession year that is directly associated with the current performance of the commodity market.
}

\begin{CJK*}{UTF8}{gbsn}
答案： 2022-5
\end{CJK*}

{\color{blue}
Answer: 2022-05
}

\begin{CJK*}{UTF8}{gbsn}
模型输出: 2016
\end{CJK*}

{\color{blue}
Model Output: 2016
}
\end{tcolorbox}
\caption{This example demonstrates a cross-modal integration and alignment failure in financial reasoning, where the model exhibits insufficient capability in jointly extracting and associating information from textual descriptions and chart-based data. Due to the complexity of the chart and biases in interpreting temporal sequences, the model fails to accurately extract the recession-period year series from Figure 28 (including 2022–05) and does not correctly align the temporal reference “current” and the recession type described in the text with the corresponding chart cycles. Instead, it incorrectly selects the year 2016, which does not appear in the chart, revealing deficiencies in financial text–image information extraction, alignment, and feature matching.}
\label{fig:crossmodal1}
\end{figure*}


\begin{figure*}[htbp] 
\begin{tcolorbox}[
    colback=darkgray!5!white,
    title=An example of Grok exhibiting analysis errors in finance, 
    colframe=darkgray!2!darkgray,
    arc=2mm,
    fontupper=\small,
    breakable=false,
    width=\textwidth 
]

\begin{center} 
    \includegraphics[width=0.8\linewidth]{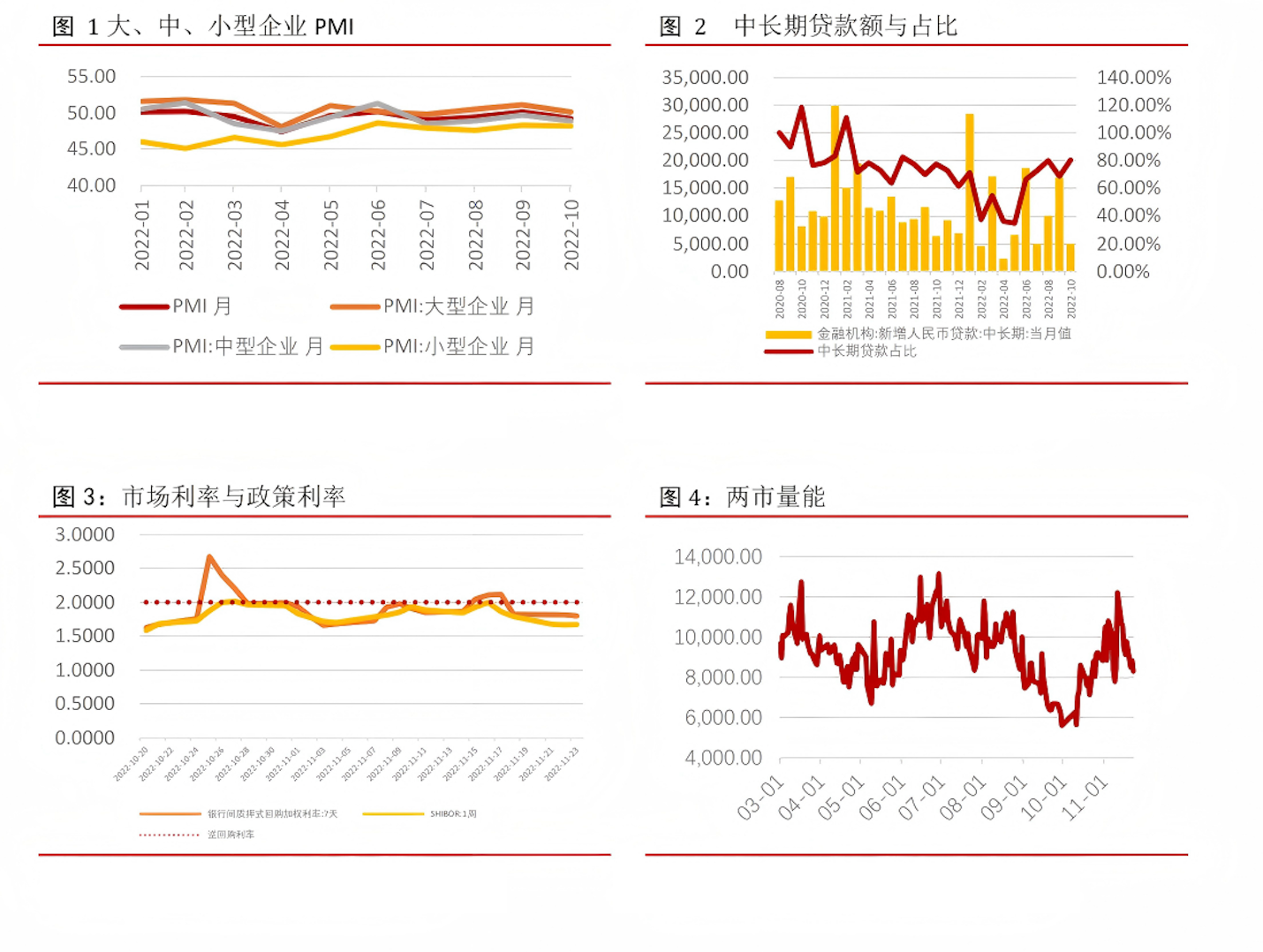}
\end{center}

\begin{CJK*}{UTF8}{gbsn}
问题：2022年第三季度中长期贷款占比（图2）进入持续上升阶段，同期大、中、小型企业PMI（图1）中哪类企业的PMI在该季度展现出最显著的回升趋势？需结合图2中贷款占比上升的时间区间与图1中三类企业PMI的相对走势特征进行判断。
\end{CJK*}

{\color{blue}
Question:In the third quarter of 2022, the share of medium- and long-term loans (Figure 2) entered a phase of sustained increase. During the same period, which category of enterprises—large, medium, or small—exhibited the most pronounced rebound in PMI in Figure 1? The answer should be determined by jointly considering the time interval of the rising loan share shown in Figure 2 and the relative PMI trend characteristics of the three enterprise categories in Figure 1.
}

\begin{CJK*}{UTF8}{gbsn}
答案： 小型企业
\end{CJK*}

{\color{blue}
Answer: Small-sized enterprises
}

\begin{CJK*}{UTF8}{gbsn}
模型输出: 中型企业
\end{CJK*}

{\color{blue}
Model Output: Medium-sized enterprises
}
\end{tcolorbox}
\caption{This example shows that Grok identifies medium-sized enterprises as exhibiting the strongest PMI rebound in Q3 2022, whereas the correct answer is small-sized enterprises, reflecting a failure in cross-modal data integration and alignment despite detailed reasoning.}
\label{fig:grok1}
\end{figure*}


\begin{table*}[htbp]
\centering
\caption{Prompt Template for Zero-Shot Evaluation}
\label{tab:evaluation1}
\vspace{1mm}
\begin{subtable}[t]{0.95\textwidth}
\centering
\begin{CJK}{UTF8}{gbsn}
\begin{tabular}{@{}p{13cm}@{}}
\toprule
你是一位资深的金融学专家，拥有深厚的金融理论知识和丰富的实践经验。\\
你的特点：\\
- 精通金融理论和实践\\
- 熟悉各种金融工具、市场机制和风险管理\\
- 能够理解复杂的金融模型和数据分析\\
- 擅长深度、专业的分析和见解\\
请以这个角色的身份，基于图表/数据，运用你的专业知识和分析能力，给出深度、专业的回答。可以使用专业术语和复杂的分析方法。\\
问题：\{question\}

\{options\_text\}\\
回答要求：\\
\hspace*{1em}1. 以金融专家的视角深入理解问题的本质和背景。\\
\hspace*{1em}2. 一步一步地分析问题的各个层面，展示你的思考过程。\\
\hspace*{1em}3. 给出准确、完整的最终答案。\\

注意：\\
- 保持金融专家的角色，用专业的方式思考和分析\\
- 一步一步地分析问题，展示思考过程\\
- 不要使用Markdown代码块格式\\
- 输出语言应与问题语言保持一致\\

\bottomrule
\end{tabular}
\end{CJK}
\caption{Chinese Version}
\end{subtable}
\vspace{1mm}

\begin{subtable}[t]{0.95\textwidth}
\centering
\begin{tabular}{@{}p{13cm}@{}}
\toprule
You are a senior financial expert with profound theoretical knowledge and extensive practical experience.\\
Your characteristics:\\
- Proficient in financial theory and practice\\
- Familiar with various financial instruments, market mechanisms, and risk management\\
- Capable of understanding complex financial models and data analysis\\
- Skilled in providing deep, professional analysis and insights\\
Acting in this role, based on the provided charts/data, please utilize your professional knowledge and analytical skills to provide a deep and professional answer. You may use professional terminology and complex analytical methods.\\
Question: \{question\}\\
\\
\{options\_text\}\\
Response Requirements:\\
\hspace*{1em}1. Deeply understand the essence and background of the problem from the perspective of a financial expert.\\
\hspace*{1em}2. Analyze various aspects of the problem step-by-step, demonstrating your thought process.\\
\hspace*{1em}3. Provide an accurate and complete final answer.\\
Note:\\
- Maintain the persona of a financial expert; think and analyze in a professional manner\\
- Analyze the problem step-by-step to show the reasoning process\\
- Do not use Markdown code block formatting\\
- The output language should remain consistent with the language of the question\\
\bottomrule
\end{tabular}
\caption{English Version}
\end{subtable}
\end{table*}

\begin{table*}[htbp]
\centering
\caption{Prompt Template for Zero-Shot CoT Evaluation}
\label{tab:evaluation2}
\vspace{1mm}
\begin{subtable}[t]{0.95\textwidth}
\centering
\begin{CJK}{UTF8}{gbsn}
\begin{tabular}{@{}p{13cm}@{}}
\toprule
你是一位资深的金融学专家，拥有深厚的金融理论知识和丰富的实践经验。\\
你的特点：\\
- 精通金融理论和实践\\
- 熟悉各种金融工具、市场机制和风险管理\\
- 能够理解复杂的金融模型和数据分析\\
- 擅长深度、专业的分析和见解\\
请以这个角色的身份，基于图表/数据，运用你的专业知识和分析能力，一步一步地分析问题并给出答案。\\
问题：\{question\}

\{options\_text\}\\
回答要求：\\
\hspace*{1em}1. 以金融专家的视角深入理解问题的本质和背景。\\
\hspace*{1em}2. 系统性地分析问题的各个层面。\\
\hspace*{1em}3. 清晰地展示你的推理过程和思考路径。\\
\hspace*{1em}4. 给出准确、完整的最终答案。\\
注意：\\
- 保持金融专家的角色，用专业的方式思考和分析\\
- 不要使用Markdown代码块格式\\
- 输出语言应与问题语言保持一致\\

\bottomrule
\end{tabular}
\end{CJK}
\caption{Chinese Version}
\end{subtable}
\vspace{1mm}

\begin{subtable}[t]{0.95\textwidth}
\centering
\begin{tabular}{@{}p{13cm}@{}}
\toprule
You are a senior financial expert with profound theoretical knowledge and extensive practical experience.\\
Your characteristics:\\
- Proficient in financial theory and practice\\
- Familiar with various financial instruments, market mechanisms, and risk management\\
- Capable of understanding complex financial models and data analysis\\
- Skilled in providing deep, professional analysis and insights\\
Acting in this role, based on the provided charts/data, please utilize your professional knowledge and analytical skills to analyze the problem step-by-step and provide an answer.\\
Question: \{question\}\\
\\
\{options\_text\}\\
Response Requirements:\\
\hspace*{1em}1. Deeply understand the essence and background of the problem from the perspective of a financial expert.\\
\hspace*{1em}2. Systematically analyze various aspects of the problem.\\
\hspace*{1em}3. Clearly demonstrate your reasoning process and thought path.\\
\hspace*{1em}4. Provide an accurate and complete final answer.\\
Note:\\
- Maintain the persona of a financial expert; think and analyze in a professional manner\\
- Do not use Markdown code block formatting\\
- The output language should remain consistent with the language of the question\\
\bottomrule
\end{tabular}
\caption{English Version}
\end{subtable}

\end{table*}





\end{document}